%% file: main.tex
\input{preambule.tex}
\begin{document}
\input{title}
%\newpage
\input{Introduction}
%\newpage
\input{Contributions}
%\newpage
\input{Resultats}

\input{Simulation} 
%\newpage
\input{Conclusion}

\newpage
\bibliography{References.bib}
\input{AnnexeV}

\end{document}

%% file: preambule.tex
\documentclass[12pt]{article}
\usepackage{pstricks}
\usepackage{pst-tree}
\usepackage[]{graphicx}
\usepackage[]{color}
\usepackage[T1]{fontenc}
\usepackage[utf8]{inputenc}
\usepackage{geometry}
\usepackage[numbers]{natbib}
\usepackage[nottoc]{tocbibind}
\usepackage{float}
\usepackage{soul}
\usepackage{amsmath}
\usepackage{amssymb}
\usepackage{setspace}
\usepackage{subcaption}
\usepackage{amsfonts}
\usepackage{dsfont}
\usepackage[frenchb]{babel}
\usepackage{caption}
\addto\captionsfrench{%
}
\usepackage[labelfont = {bf}, 
  textfont ={it}, 
  margin =1cm,
  font=small, 
  tablename = Tableau ]{caption}
  
\usepackage[labelfont = {bf}, 
  textfont ={it}, 
  margin =1cm,
  font=small, 
  figurename = Figure ]{caption}
  
\addto\captionsfrench{
  
}

\usepackage{amsthm} 
\usepackage{bbm}
\usepackage{booktabs}
\usepackage{float}
\usepackage{colortbl}
\usepackage{dcolumn}
\usepackage{icomma}
\usepackage{fancyhdr}
\usepackage{titlesec}
\titleformat{\section}
{\Huge\bfseries}{\thesection}{12pt}{}%[\titlerule]
\titleformat{\subsection}
{\large\bfseries}{\thesubsection}{12pt}{}[]
\titleformat{\subsubsection}
{\bfseries}{\thesubsubsection}{12pt}{}[]
\usepackage{listings}
\usepackage[flushleft]{threeparttable}
\usepackage{caption}
\usepackage[tikz]{bclogo}
\usepackage{pgfornament}
\usepackage{url} 
\usepackage[hidelinks]{hyperref}
\usepackage{fancyhdr} 
\usepackage{fancybox}
\usepackage{pifont}
\usepackage{multirow}
\usepackage{booktabs}
\usepackage{makecell}

%% file: title.tex
\begin{titlepage}
\newcommand{\HRule}{\rule{\linewidth}{0.5mm}} 
\begin{center}
 \textbf{A copula-based competing risks model with application to cancer risk estimation in family studies}   
\end{center}

\vspace{0.5cm}
\begin{center}
   Siriac Seboka\textsuperscript{1}, Laurent Briollais\textsuperscript{2,3}, Yun-Hee Choi\textsuperscript{4}, on behalf of BCFR,
Lajmi Lakal-Chaieb \textsuperscript{1,*}\\

 {\small \textsuperscript{1}Département de Mathématiques et de Statistique, Université Laval, Québec, Canada}\\{\small \textsuperscript{2}Lunenfeld-Tanenbaum Research Institute, Mount Sinai Hospital, Toronto, Canada}\\
 {\small \textsuperscript{3}Biostatistics division, Dalla Lana School of Public Health, University of Toronto, Toronto, Canada}\\
 {\small \textsuperscript{4}Department of Epidemiology and Biostatistics, Western University, London, Canada}\\
{\small \textsuperscript{*}Correspondence: Lajmi Lakal-Chaieb (lakhal@mat.ulaval.ca)} 
\end{center}

\noindent \textbf{Abstract}:
Cancer risk estimation tools are used by clinicians to evaluate the risk of individuals who carry high-risk pathogenic variants. In this article, we propose a new competing risks model designed to provide personalized estimates of the risk of developing breast or ovarian cancer as the first event among women carrying pathogenic variants in the \textit{BRCA1} gene. Our approach incorporates both the personal and complete family history of cancer of these high-risk women and introduces some flexibly in the modeling of the dependence among members of the same family. We use a Gaussian copula with a correlation matrix expressed in terms of the kinship coefficients of pairs of members. Furthermore, we develop an iterative estimation algorithm to estimate the parameters of the underlying model in the presence of a selection bias inherent to the family recruitment process. The performance of the proposed method is evaluated through a simulation study. An application to the Breast Cancer Family Registry data highlights the ability of the model to provide personalized estimation of the risk of developing breast cancer or ovarian cancer as the first event within the competing risks framework.

\vspace{0.2cm}


\end{titlepage}                             

%% file: Introduction.tex
\section{Introduction}

\noindent The clinical management of women who carry pathogenic variants in the \textit{BRCA1} and \textit{BRCA2} predisposition genes is based on an accurate and individualized assessment of their cancer risk. These women face high lifetime risks, reaching approximately 72\% for breast cancer (BC) and 44\% for ovarian cancer (OC) by age 80 among \textit{BRCA1} carriers, and 69\% and 17\%, respectively, among \textit{BRCA2} carriers \citep {kuchenbaecker2017risks}. To reduce these risks and to prevent cancers to occur in these high-risk women, several options such as risk-reducing bilateral mastectomy, risk-reducing bilateral salpingo-oophorectomy \citep{larsen2014hereditary} or regular mammographic surveillance are often proposed. However, the choice of mammographic screening or risk-reducing interventions depends on a reliable estimate of the absolute risk of both BC and OC but also on the temporal dynamics according to which the two cancers may occur.

\noindent In this context, the analysis of the first tumor event defined here as the occurrence of BC or OC is important. For a woman who has not yet developed cancer at the time of evaluation, the clinical decision depends both on the expected time to the first event and on the probability that this first event will be BC or OC. This issue falls within the framework of competing risks. The occurrence of a first cancer alters the history, subsequent surveillance, and interpretation of the risk associated with the other cancer.
Current prevention protocols for women at high genetic risk rely heavily on validated clinical prediction models, such as BOADICEA (integrated into the CanRisk tool), BRCAPRO, or the Cuzick–Tyrer model \citep{antoniou2004boadicea,lee2019boadicea,carver2021canrisk, parmigiani1998determining,tyrer2004breast}%\textbf{refs?}
. These tools play an essential role in estimating individual risk and in the clinical decision-making. However, they are not directly designed to model, within a selected family context, the joint distribution of the time to the first tumor event and the type of cancer observed first. In practice, the separate (or marginal) assessment of BC and OC risks leads to an incomplete representation of individual trajectories, as both cancers share significant biological and genetic links \citep{angeli2020predisposition}.

\noindent This difficulty is exacerbated in family studies, where observations cannot be considered independent. Members of the same family share part of their genetic makeup as well as certain unobserved environmental or behavioral factors. The resulting intra-family dependence should therefore not be assumed uniform across all relatives, but rather modulated by their degree of biological kinship. Furthermore, family registries, such as the Breast Cancer Family Registry (BCFR) \cite{john2004breast}%\textbf{ref?}
, often recruit families based on an affected proband. This sampling mechanism introduces an ascertainment bias which, if not accounted for, can lead to an overestimation of cumulative incidences and absolute risks of cancer \citep{ranola2019exploring}.

\noindent During the last couple of decades, several authors have examined the modeling and measures of dependence for multivariate competing risks; see \citep{su2020association} and the references therein. However, most common approaches rely on the cause-specific or subdistribution models combined with shared random effects, often referred to as flexible fragility models \citep{oakes1989bivariate,clayton1985multivariate,katsahian2006analysing,scheike2010semiparametric}%[\textcolor{red}{REF : citer quelques references parmi les references 6 a 20 du papier de statistics in medicine 2020}]
. These approaches allow for intra-family correlation, but they generaly impose a restrictive symmetric dependence structure. In particular, a shared fragility common to all members of a family tends to induce homogeneous dependence among pairs of relatives, without explicitly representing the expected differences between, for example, two sisters, a mother and her daughter, or two more distant-related individuals. Moreover, the inference methodology associated to these models can not be directly applied in the presence of a selection bias. Recently, \cite{lee2026flexible} considered a fraily-based approach for multivariate competing risks in the presence of a selection bias. Unfortunately, this model does not allow for a rich variety of expressions of dependence measures and risk functions. 

\noindent To address these limitations, we propose here a copula-based risk prediction approach for analyzing BC and OC as competing risks in families carrying pathogenic variants in \textit{BRCA1}. The model combines marginal cause-specific hazard functions to describe the failure processes specific to each cancer, and a Gaussian copula to represent the dependence between times to the first event among pairs of members of the same family. The copula’s correlation matrix is indexed by biological kinship coefficients and by three family dependence or heritability parameters, denoted $h_{11}$, $h_{12}$, and $h_{22}$, associated respectively with the dependence between BC events, between BC and OC events, and between OC events. To correct for the ascertainment bias introduced by BCFR recruitment protocol, we construct a conditional likelihood that takes into account information about the proband from whom the family is selected at the time of recruitment.

\noindent From a computational standpoint, the joint estimation of marginal and dependence parameters of the considered model can be computationally intensive. In this work, we develop an iterative algorithm to estimate the parameters of our model, which we refer to as doubly \textrm{EM}-algorithm. We treat the unobserved event indicators for the censored observations as missing data. The doubly \textrm{EM}-algorithm iterates between computing the expected values of the missing data and maximizing two log-likelihood functions: one for the marginal and the other for the dependence parameters. This strategy reduces computational complexity while maintaining a flexible modeling of the family dependency.

\noindent The remainder of the article is organized as follows. Section 2 describes the structure of the data from BCFR. Section 3 presents the copula-based competing risks model as well as some common properties of copulas. Section 4 details the proposed estimation procedure. Section 5 applies the proposed method to data from families carrying pathogenic variants in \textit{BRCA1}. Section 6 presents a simulation study to validate the proposed model. We conclude with a summary in Section 7.

%% file: Contributions.tex
\section{Data Set}

\noindent
The data used in this study are from the Breast Cancer Family Registry (BCFR), a multi-center family-based registry established in 1995 and including participants from the United States, Canada, and Australia \citep{john2004breast}. Families were ascertained through a proband and were enriched for familial breast and/or ovarian cancer. The registry collected detailed information on cancer diagnoses, age at diagnosis, risk-reducing surgeries, mammographic screening history, family relationships, and pathogenic variant status in \textit{BRCA1} and \textit{BRCA2} genes through baseline and follow-up 
questionnaires \citep{terry2016cohort}.

\noindent For the present analysis, we focused on families known to carry a pathogenic variant in the \textit{BRCA1} gene. Each individual was assumed to enter follow-up at age 18 and was followed-up until the first occurrence of breast cancer, ovarian cancer, or censored at the age of last follow-up. Thus, breast and ovarian cancers were treated as competing first-event outcomes. The family-based sampling design, in which families were identified through an affected proband, was explicitly accounted for in the likelihood construction to reduce ascertainment bias.

\noindent
All participating centers obtained written informed consent from study 
participants, and the study protocols were approved by the relevant local ethical review committees.

\section{Notation and Model}

\noindent We consider a dataset of \( I \) independent families. For  \( i = 1, \dots, I \), we assume that family \( i \) is composed of \( n_i + 1 \) members, indexed by \( j=0,\cdots,n_i \). The index $j=0$ corresponds to the proband. Adopting the notation of cause-specific hazards, we let the survival data for individual $j$ in family $i$ be $(T_{ij},\varepsilon_{ij})$ where $T_{ij}$ is the age at occurence of the first event and $\varepsilon_{ij} \in \{1,2\}$ is a binary variable equal to 1 if $T_{ij}$ corresponds to the age at onset of breast cancer and 2 if it corresponds to the age at onset of ovarian cancer. Additionnaly, one also observes the covariate $X_{ij}$, which is a binary variable indicating gene carrier status (1 for carriers and 0 otherwise). The study includes a total of \( N = I + \sum_{i=1}^{I} n_i \) participants, i.e., $I$ probands and their relatives.

\subsection{Marginal distribution}

\noindent We specify the marginal cause-specific hazard function for a type $l$ event by
$$\lambda_l(t|X_{ij}) = \lim\limits_{\Delta t \rightarrow 0} \frac{1}{\Delta t} P(t < T_{ij} \le t + \Delta t,  \varepsilon_{ij}=l|T_{ij} > t, X_{ij} ); \, \, \, t>0, \, \, \, l=1,2$$

\noindent and we assume the Cox model

 \[ \lambda_l(t|X_{ij}) = \lambda_{0l}(t) \exp\left\{\beta_l X_{ij} \right\}, \]  where $\lambda_{0l}(t)$ is the baseline hazard function and $\beta_l$ the regression coefficient for event $l$.

 \noindent Without loss of generality, we consider a parametric approach and assume a baseline hazard function
 $\lambda_{0l}(t) =\alpha_l \lambda_l t^{\alpha_l-1}$ specified according to the Weibull distribution with shape and scale parameters equal to $\alpha_l$ and $\lambda_l$, respectively.
 
 \noindent The hazard and survival functions of \( T \) are defined, respectively, by 
$$\lambda_T(t|X_{ij})=\lambda_1(t|X_{ij})+\lambda_2(t|X_{ij})$$
and

$$S_T(t|X_{ij})=P(T>t|X_{ij})=\exp \left\{-\int_{0}^{t}\lambda_T(u|X_{ij})du\right\}.$$ 
The cumulative incidence function for event \( l \) is 
$$F_l(t|X_{ij})=P(T\leq t, \varepsilon_{ij}=l|X_{ij})=\int_{0}^{t}\lambda_l(u|X_{ij})S_T(u|X_{ij})du.$$ 
Therefore, one has $$P(\varepsilon_{ij}=l|X_{ij})=F_l(\infty|X_{ij})= \int_{0}^{\infty}\lambda_l(u|X_{ij})S_T(u|X_{ij})du$$ and $$S_l(t|X_{ij})=1 - P(T \le t|\varepsilon_{ij}=l,X_{ij})=1-\dfrac{F_l(t|X_{ij})}{F_l(\infty|X_{ij})}.$$
Finally, the conditional cause-specific density function is
$$f_l(t|X_{ij}) = -\frac{\partial S_l(t|X_{ij})}{\partial t} = \frac{\lambda_l(t|X_{ij}) S_T(t|X_{ij})}{F_l(\infty|X_{ij})}.$$

\noindent The marginal parameters are then $\theta=(\lambda_1,\alpha_1,\beta_1,\lambda_2,\alpha_2,\beta_2)^\top$.

\subsection{Dependence model}

%%%%%%%%%%%%%%%%%%%%%%%%%%%%%%%%%%%%%%%%%%%%%%%%%%%%%
\noindent We model the dependencies between the survival data in family $i$ in two steps. In the first step, we assume that the event indicators $\{\varepsilon_{i0}, \varepsilon_{i1}, \cdots, \varepsilon_{in_i}\}$ are independent given the covariates. In other terms, we assume that
$$P(\varepsilon_{i0}=l_{i0}, \varepsilon_{i1}=l_{i1}, \cdots, \varepsilon_{in_i}=l_{in_i}| X_{i0},\cdots,X_{in_i}) = \prod_{j=0}^{n_i} P(\varepsilon_{ij}=l_{ij}|X_{ij})$$
for all $\{l_{i0},l_{i1},\cdots,l_{in_i}\} \in \{1,2\}^{n_i}.$ In the second step, we use a Gaussian copula to model the dependencies between the event times $\{T_{i0},T_{i1},\cdots,T_{in_i}\}$ given the event types $\{\varepsilon_{i0}=l_{i0}, \varepsilon_{i1}=l_{i1}, \cdots, \varepsilon_{in_i}=l_{in_i}\}$. Under such an assumption, one has
$$P(T_{i0}>t_{i0},\dots,T_{in_i}>t_{in_i}|\varepsilon_{i1}=l_1, \dots, \varepsilon_{in_i}=l_{in_i})=\mathcal{D}_{v_i} \left\{ S_{l_0}( t_{i0}|Z_{i0}), \dots , S_{l_{in_i}}( t_{in_i}|Z_{in_i})\right\},$$ where \( \mathcal{D}_{v_i} \) is a Gaussian copula indexed by the correlation matrix \( v_i \). The diagonal entries of $v_i$ are equal to 1. The off-diagonal entries $(j,k); \, \, \, 0 \le j \neq k \le n_i$ of $v_i$ are equal to $h_{l_j,l_k}\times \phi^{(i)}_{jk}$, where $h_{11}, h_{12}, h_{21}$ and $h_{22}$ are the heritability parameters and 
$\phi^{(i)}_{jk}$ is the kinship coefficient, which is the probability that two randomly selected alleles from the same locus of relatives $j$ and $k$ in family $i$ are identical by descent.

\noindent The heritability parameter $h_{l_j,l_k} 
\in [0,1]$ measures the strength of the association between the times of event types $l_j$ and $l_k$. For instance, $h_{11}$ assesses the dependence in the pair $(T_{ij},T_{ik})$ when $(\varepsilon_{ij},\varepsilon_{ik})=(1,1)$, whereas $h_{12}=h_{21}$ measures the correlation in the pair $(T_{ij},T_{ik})$ when $(\varepsilon_{ij},\varepsilon_{ik}) \in \left\{(1,2),(2,1)\right\}.$ On the other hand, $\phi^{(i)}_{jk} \in [0,1]¸$ indicates the degree of relatedness between indivudals $j$ and $k$ in family $i$. For instance, $\phi^{(i)}_{jk}=1/4$ if $j$ and $k$ are sisters or mother and daughter, $\phi^{(i)}_{jk}=1/8$ if $j$ and $k$ are niece and aunt, and $\phi^{(i)}_{jk}=1/16$ if $j$ and $k$ are first cousins. 

\noindent Consider for instance a family $i$ composed of $n_i+1=4$ members such that $\varepsilon_{i0}=1,\varepsilon_{i1}=1, \varepsilon_{i2}=2, \varepsilon_{i3}=2.$ The correlation matrix is 
\begin{equation*}
    v_i= \begin{pmatrix}1&h_{11}\times\phi_{01}^{(i)} & h_{12}\times\phi_{02}^{(i)}  & h_{12}\times\phi_{03}^{(i)}\\
    h_{11}\times\phi_{10}^{(i)} & 1 &  h_{12}\times\phi_{12}^{(i)} &  h_{12}\times\phi_{13}^{(i)}\\ 
    h_{21}\times\phi_{20}^{(i)} &  h_{21}\times\phi_{21}^{(i)} & 1 &  h_{22}\times\phi_{23}^{(i)}\\
     h_{21}\times\phi_{30}^{(i)} &  h_{21}\times\phi_{31}^{(i)} &  h_{22}\times\phi_{32}^{(i)} & 1  \end{pmatrix}.
\end{equation*}

\noindent The dependence parameters are then $h=(h_{11},h_{12},h_{22})^\top$.

\subsection{Useful quantities of interest}

\noindent In this section, we derive the conditional distributions associated with the event times of family members, including both individual times $T_{ij}$ and joint times $(T_{ij},T_{ik})$ for the first observed event, conditional on the available information from the affected proband \textbf{(j=0)}. These quantities form the basis for the likelihood contributions developed in the subsequent sections.

\noindent For $i=1,\cdots,I$, $j=1,\cdots,n_i$, one has
$$P(T_{i0}>t_{i0}, T_{ij}>t_{ij}|\varepsilon_{i0}=l_{i0}, \varepsilon_{ij}=l_{ij}, X_{i0},X_{ij}) = \mathcal{D}\{S_{l_{i0}}(t_{i0}|X_{i0}),S_{l_{ij}}(t_{ij}|X_{ij})\}.$$
For simplicity of notation, we will omit to indicate the dimension and the correlation matrices of the Gaussian copulas from here on. This information can easily be deduced from the context. For instance, in the equality above, the dimension of the Gaussian copula is 2 and the correlation matrix is 
$$\begin{pmatrix} 1 & h_{l_{i0}l_{ij}}\times \phi_{0j}^{(i)} \\ h_{l_{i0}l_{ij}}\times \phi_{0j}^{(i)} & 1\end{pmatrix}.$$
Therefore the cause-specific conditional survival and density functions of $T_{ij}$ given $T_{i0}=t_{i0}$ are respectively equal to
$$P(T_{ij}>t_{ij}, \varepsilon_{ij}=l_{ij}|T_{i0}=t_{i0}, \varepsilon_{i0}=l_{i0},  X_{ij},X_{i0}) = \mathcal{D}^{10}\{S_{l_{i0}}(t_{i0}|X_{i0}),S_{l_{ij}}(t_{ij}|X_{ij})\}F_{l_{ij}}(\infty|X_{ij})$$
and
\begin{align*}
  &-\frac{\partial}{\partial t}P(T_{ij}>t,  \varepsilon_{ij}=l_{ij}|T_{i0}=t_{i0}, \varepsilon_{i0}=l_{i0}, X_{ij},X_{i0})|_{t=t_{ij}} \\
  &= \mathcal{D}^{11}\{S_{l_{i0}}(t_{i0}|X_{i0}),S_{l_{ij}}(t_{ij}|X_{ij})\} f_{l_{ij}}(t_{ij}|X_{ij}),  
\end{align*}
where $\mathcal{D}^{10}(u,v)=\partial \mathcal{D}(u,v)/\partial u$ and $\mathcal{D}^{11}(u,v)=\partial^2 D(u,v)/\partial u \partial v$.

\noindent Moreover, for $i=1,\cdots,I$, $1 \le j < k \le n_i$, one has
\begin{align*}
  &P(T_{i0}>t_{i0},T_{ij}>t_{ij}, T_{ik}>t_{ik}|\varepsilon_{i0}=l_{i0}, \varepsilon_{ij}=l_{ij}, \varepsilon_{ik}=l_{ik}, X_{i0},X_{ij},X_{ik}) \\ 
  &= \mathcal{D}\{S_{l_{i0}}(t_{i0}|X_{i0}), S_{l_{ij}}(t_{ij}|X_{ij}), S_{l_{ik}}(t_{ik}|X_{ik})\}.  
\end{align*}

\noindent Therefore, the joint distribution of $(T_{ij},T_{ik})$ given $T_{i0}=t_{i0}$ is characterized by
\begin{align*}
&P(T_{ij}>t_{ij}, T_{ik}>t_{ik},  \varepsilon_{ij}=l_{ij},  \varepsilon_{ik}=l_{ik}|T_{i0}=t_{i0}, \varepsilon_{i0}=l_{i0}, X_{i0}, X_{ij}, X_{ik}) \\
&= \mathcal{D}^{100}\{S_{l_{i0}}(t_{i0}|X_{i0}), S_{l_{ij}}(t_{ij}|X_{ij}), S_{l_{ik}}(t_{ik}|X_{ik})\}F_{l_{ij}}(\infty|X_{ij})F_{l_{ik}}(\infty|X_{ik}),
\end{align*}
\begin{align*}
 &-\frac{\partial}{\partial t} P(T_{ij}>t, T_{ik}>t_{ik},  \varepsilon_{ij}=l_{ij}, \varepsilon_{ik}=l_{ik}|T_{i0}=t_{i0}, \varepsilon_{i0}=l_{i0},X_{i0}, X_{ij}, X_{ik})|_{t=t_{ij}}\\
 &= \mathcal{D}^{110}\{S_{l_{i0}}(t_{i0}|X_{i0}), S_{l_{ij}}(t_{ij}|X_{ij}), S_{l_{ik}}(t_{ik}|X_{ik})\} f_{l_{ij}}(t_{ij}|X_{ij})F_{l_{ik}}(\infty|X_{ik}),
\end{align*}
and
\begin{align*}
  &\frac{\partial^2}{\partial t \partial s} P(T_{ij}>t, T_{ik}>s, \varepsilon_{ij}=l_{ij}, \varepsilon_{ik}=l_{ik}|T_{i0}=t_{i0}, \varepsilon_{i0}=l_{i0},X_{i0}, X_{ij}, X_{ik})|_{t=t_{ij},s=t_{ik}}\\
  &= \mathcal{D}^{111}\{S_{l_{i0}}(t_{i0}|X_{i0}),S_{l_{ij}}(t_{ij}|X_{ij}),S_{l_{ik}}(t_{ik}|X_{ik})\} f_{l_{ij}}(t_{ij}|X_{ij})f_{l_{ik}}(t_{ik}|X_{ik}),  
\end{align*}
where $\mathcal{D}^{100}(u,v,w)=\partial \mathcal{D}(u,v,w)/\partial u$, $D^{110}(u,v,w)=\partial^2 \mathcal{D}(u,v,w)/\partial u \partial v$, and \\ $\mathcal{D}^{111}(u,v,w)=\partial^3 \mathcal{D}(u,v,w)/\partial u \partial v \partial w$.

\section{Inference procedure}

\subsection{Observed and complete data}
\noindent For \(i=1, \dots, I\), let \(a_i\) be the examination age of the proband in family \(i\). According to the data collection protocol, a family \(i\) is included in the data set if and only if the proband has observed either breast or ovarian cancer prior to the examination age. In our notation, $j=0$ corresponds to the proband. Thus, the inclusion condition for family \(i\) is \(\{T_{i0} \leq a_i\}\).

\noindent The survival data \((T_{ij},\varepsilon_{ij})\) for the non-probands are not always observed due to right-censoring by age at last follow-up. For \(i=1, \dots, I; j=0, \dots, n_i\), we only observe $(\tilde T_{ij},\delta_{ij})$, where \(\tilde{T}_{ij} = \min(T_{ij}, C_{ij})\), $C_{ij}$ is the age at last follow-up for individual $j$ in family $i$ and \(\delta_{ij} = 1_{\{T_{ij} < C_{ij}\}} \times \varepsilon_{ij} \in \{0, 1, 2\}\). We assume that $C_{ij}$  is independent of \(T_{ij}\) conditionally on \(X_{ij}\). Thus, the observed data is 
\[
{\cal O} = \left\{ ( a_i, \tilde{T}_{ij}, \delta_{ij}, X_{ij} ) ; j =0, \dots, n_i, \, i = 1, \dots, I \right\}.
\]

\noindent When $\delta_{ij}=0$, we do not observe $\varepsilon_{ij}$. In the next section, we develop what we refer to as a doubly-EM-algorithm to estimate the parameters of our model. This algorithm requires the specification of a complete data set, which we define as
$${\cal C} = \left\{ ( a_i, \tilde{T}_{ij}, \delta_{ij}, \varepsilon_{ij}, X_{ij} ) ; j =0, \dots, n_i, \, i = 1, \dots, I \right\}.$$

\noindent In other words, the complete data include the event types for the censored observations. Finally, note that $\delta_{ij}=\varepsilon_{ij}$ when $\delta_{ij}>0$.

\subsection{Parameter estimation with complete data}
\noindent The set of parameters to be estimated from the data is \(\Theta = (\theta^\top, h^\top)^\top\). We begin by describing an iterative algorithm to estimate $\Theta$ from the complete dataset ${\cal C}$. 

\noindent We consider two conditional log-likelihood functions \(l^{(1)}(\theta \mid h, {\cal C})\) and \(l^{(2)}(h \mid \theta, {\cal C})\), referred to as univariate and bivariate, respectively. The spirit of these functions comes from the two-stage procedure, where the marginal parameters are estimated by maximizing a univariate log-likelihood function first and the association parameter is obtained by maximizing a bivariate log-likelihood function, given the estimates of the marginal parameters, afterwards \citep{andersen2005two,su2020association}. In our context, the univariate conditional likelihood function depends, not only on $\theta$, but also on $h$ due to the ascertainment correction. Therefore, it is impossible to apply the two-stage procedure directly. We generalize the latter procedure by proposing an algorithm that iterates between the maximizations of \(l^{(1)}(\theta \mid h, {\cal C})\) and \(l^{(2)}(h \mid \theta, {\cal C})\), until convergence. 

\subsubsection{Univariate log-likelihood function for complete data}
\noindent The univariate likelihood for complete datra is the sum of the individual contributions of probands and non-probands:
\begin{equation*}
    l^{(1)}(\theta |h,{\cal C}) = \sum_{i=1}^{I} l^{(1)}_i(\theta |h,{\cal C}) = \sum_{i=1}^{I} \sum_{j=0}^{n_i} l^{(1)}_{ij}(\theta |h,{\cal C}) = \sum_{i=1}^{I} \sum_{j=0}^{n_i} \log \left[ L^{(1)}_{ij}(\theta |h,{\cal C}) \right].
\end{equation*}

\noindent The contribution of a proband who has observed either breast or ovarian cancer prior to the entry to the study is:
\begin{equation*}
    L^{(1)}_{i0}(\theta |h,{\cal C}) = -\frac{\partial P(T_{i0} > t, \varepsilon_{i0} = \delta_{i0} | T_{i0} < a_i)}{\partial t}|_{t=t_{io}} =  \frac{f_{\delta_{i0}}(t_{i0}|X_{i0})}{1 - S_T(a_i|X_{i0})}.
\end{equation*}
On the other hand, the contribution of a proband who was diagnosed with either breast or ovarian cancer at the recruitement (\(T_{i0} = a_i\)) to the study is 
\begin{equation*}
    L^{(1)}_{i0}(\theta |h,{\cal C}) = -\frac{\partial P(T_{i0} > t, \varepsilon_{i0} = \delta_{i0} )}{\partial t}|_{t=t_{io}} = f_{\delta_{i0}}(t_{i0}).
\end{equation*}

\noindent Consider now a non-proband $(i, j)$ who have observed either breast or ovarian cancer at age $t_{ij}$. Its contribution to the univariate likelihood function is

\begin{align*}
    L^{(1)}_{ij}(\theta |h,{\cal C}) &= -\frac{\partial P(T_{ij} > t, \varepsilon_{ij} = \delta_{ij} | T_{i0} = t_{i0}, \varepsilon_{i0} = \delta_{i0}, X_{i0}, X_{ij})}{\partial t} \\  &= D^{11}\{S_{\delta_{ij}}(t_{ij}|X_{ij}), S_{\delta_{i0}}(t_{i0}|X_{i0})\} f_{\delta_{ij}}(t_{ij}|X_{ij}).
\end{align*}

\noindent Similarly, the contribution of a non-proband individual $j$ from family $i$ with a censored observation ($\delta_{ij}=0$) is 
\begin{align*}
    L^{(1)}_{ij}(\theta |h,{\cal C}) &= \prod_{l=1}^{2} P(T_{ij} > t_{ij}, \varepsilon_{ij} = l | T_{i0} = t_{i0}, \varepsilon_{i0} = \delta_{i0})^{1_{\{\varepsilon_{ij}=l\}}} \\
    &=\prod_{l=1}^{2}\left[ D^{01}\left\{S_l(t_{ij}), S_{\delta_{i0}}(t_{i0})\right\} F_l(\infty|X_{ij})\right]^{1_{\{\varepsilon_{ij}=l\}}}.
\end{align*}

\subsubsection{Bivariate log-likelihood function for complete data}
\noindent We define the bivariate log-likelihood function as the sum of the contributions of all pairs of non-probands $(j,k); 1\le j\neq k \le n_i$ from all families $i=1,\cdots,I$.
\begin{align*}
    l^{(2)}(h|\theta,{\cal C}) = \sum_{i=1}^{I} l^{(2)}_i(h|\theta,{\cal C}) = \sum_{i=1}^{I} \sum_{1 \leq j < k \leq n_i} l^{(2)}_{ijk}(h|\theta,{\cal C})=\sum_{i=1}^{I} \sum_{1 \leq j < k \leq n_i} \log\left[L^{(2)}_{ijk}(h|\theta,{\cal C})\right].
\end{align*}

\noindent There are three figure cases according to the values of $(\delta_{ij},\delta_{ik})$.

\noindent If $\delta_{ij}>0$ and $\delta_{ik}>0$, then both $T_{ij}$ and $T_{ik}$ are observed and one has
\begin{align*}
    L^{(2)}_{ijk}(h|\theta,{\cal C}) &= \frac{\partial^2 P(T_{ij}>t, T_{ik}>s, \varepsilon_{ij}=\delta_{ij}, \varepsilon_{ik}=\delta_{ik}|T_{i0}=t_{i0}, \varepsilon_{i0}=\delta_{i0})}{\partial t \partial s}|_{t=t_{ij},s=t_{ik}} \\
    &= D^{111}\left\{S_{\delta_{ij}}(t_{ij}|X_{ij}),S_{\delta_{ik}}(t_{ik}|X_{ik}),S_{\delta_{i0}}(t_{i0}|X_{i0})\right\}f_{\delta_{ij}}(t_{ij}|X_{ij}) f_{\delta_{ik}}(t_{ik}|X_{ik}) \\
    &\propto D^{111}\left\{S_{\delta_{ij}}(t_{ij}|X_{ij}),S_{\delta_{ik}}(t_{ik}|X_{ik}),S_{\delta_{i0}}(t_{i0}|X_{i0})\right\}.
\end{align*}

\noindent If $\delta_{ij}>0$ and $\delta_{ik}=0$, then $T_{ij}$ is observed and $T_{ik}$ is censored and one has

\begin{align*}
    L^{(2)}_{ijk}(h|\theta,{\cal C}) &= \prod_{l=1}^2 \left[-\frac{\partial P(T_{ij}>t, T_{ik}>t_{ik}, \varepsilon_{ij}=\delta_{ij}, \varepsilon_{ik}=l|T_{i0}=t_{i0}, \varepsilon_{i0}=\delta_{i0})}{\partial t}|_{t=t_{ij}}\right]^{1_{\{\varepsilon_{ik}=l\}}} \\
    &\propto \prod_{l=1}^2 \left[ D^{101}\left\{S_{\delta_{ij}}(t_{ij}|X_{ij}),S_{l}(t_{ik}|X_{ik}),S_{\delta_{i0}}(t_{i0}|X_{i0})\right\}\right]^{1_{\{\varepsilon_{ik}=l\}}}.
\end{align*}

\noindent If $\delta_{ij}>0$ and $\delta_{ik}=0$, then both $T_{ij}$ and $T_{ik}$ are censored and one has

\begin{align*}
    L^{(2)}_{ijk}(h|\theta,{\cal C}) &= \prod_{l=1}^2 \prod_{l'=1}^2\left[ P(T_{ij}>t_{ij}, T_{ik}>t_{ik}, \varepsilon_{ij}=l, \varepsilon_{ik}=l'|T_{i0}=t_{i0}, \varepsilon_{i0}=\delta_{i0})\right]^{1_{\{\varepsilon_{ij}=l,\varepsilon_{ik}=l'\}}} \\
    &\propto \prod_{l=1}^2 \left[ D^{001}\left\{S_{l}(t_{ij}|X_{ij}),S_{l'}(t_{ik}|X_{ik}),S_{\delta_{i0}}(t_{i0}|X_{i0})\right\}\right]^{1_{\{\varepsilon_{ij}=l,\varepsilon_{ik}=l'\}}}.
\end{align*}

\subsubsection{Iterative algorithm}

\noindent Given the complete data, one may estimate $\Theta$ following these steps:
\begin{enumerate}
\item[{\bf (a)}] Begin with a starting value for $h$, say $h_0$
    \item[{\bf (b)}] Maximize $l^{(1)}(\theta|h_0,{\cal C})$ to obtain $\hat\theta$.
    \item[{\bf (c)}] Maximize $l^{(2)}(h|\hat\theta,{\cal C})$ to obtain $\hat h$.
    \item[{\bf (d)}] Maximize $l^{(1)}(\theta|\hat h,{\cal C})$ to obtain $\hat\theta$.
    \item[{\bf (e)}] Repeat {\bf (c)}--{\bf (d)} until convergence.
\end{enumerate}

%%%%%%%%%%%%%%%%%%%%%%%%%%%%%%%%%%%%%%%%%%%%%%%%%%%%%%%%%%%%%%%%%%%%%%%%%%%%%%%%%%%%%%%%%%%%%%%%%%%%%%%%%%%%%%%%%%%%%%%%%%%%%%%%%%%%%%%%%%%%%%%%%%%%%%%%%%%%%%%%%%%%%%%%%%%%%%%%%%%%%%%%%%%%%%

\subsection{Doubly-EM-algorithm}
\noindent When $\delta_{ij}=0$, it is not possible to observe the event type $\varepsilon_{ij}$. In this section, we propose a doubly-EM-algorithm to recover the missing information and to estimate the parameters $\Theta$ from the observed data ${\cal O}$. 

\noindent To this end, one observes that the univariate log-likelihood function for the complete data $l^{(1)}(\theta |h,{\cal C})$ is expressed linearly in terms of the missing information $1_{\{\epsilon_{ij}=l\}}$. Therefore, $E\left[l^{(1)}(\theta |h,{\cal C}) | {\cal O}\right]$ has a linear expression in terms of 
$$W_{ij}^l = E\left[1_{\{\varepsilon_{ij}=l\}}|{\cal O}\right]= P(\varepsilon_{ij}=l|T_{ij}>t_{ij},T_{i0}=t_{i0},\varepsilon_{i0}=\delta_{i0}).$$

\noindent Similarly, $l^{(2)}(h|\theta,{\cal C})$ is expressed linearly in terms of $1_{\{\varepsilon_{ij}=l\}}$ and $1_{\{\varepsilon_{ij}=l,\varepsilon_{ik}=l'\}}$. Therefore, $E\left[l^{(2)}(h |\theta,{\cal C}) | {\cal O}\right]$ has a linear expression in terms of 
$$W^{l}_{ijk} = E\left[1_{\{\varepsilon_{ij}=l\}}|{\cal O}\right]= P(\varepsilon_{ij}=l|T_{ij}>t_{ij},T_{ik}=t_{ik}, \varepsilon_{ik}=\delta_{ik},T_{i0}=t_{i0},\varepsilon_{i0}=\delta_{i0})$$
and
$$W^{ll'}_{ijk} = E\left[1_{\{\varepsilon_{ij}=l,\varepsilon_{ik}=l'\}}|{\cal O}\right]= P(\varepsilon_{ij}=l,\varepsilon_{ik}=l'|T_{ij}>t_{ij},T_{ik}>t_{ik}, T_{i0}=t_{i0},\varepsilon_{i0}=\delta_{i0}).$$

\noindent Detailed expressions of $W_{ij}^l$, $W^{l}_{ijk}$ and $W^{ll'}_{ijk}$ in terms of $\Theta$ are given in the appendices. 

\noindent Combining the iterative algorithm presented in the previous section with the above expected expressions yields the following doubly-EM-algorithm to estimate the parameters $\Theta$ based on the observed data $\cal O$. 

\begin{enumerate}
    \item[{\bf (a)}] Begin with initial values for the parameters $\Theta$.
    \item[{\bf (b)}] $E_1$-step: Compute the expected values $W_{ij}^l$.
    \item[{\bf (c)}] $M_1$-step: Maximize $E\left[l^{(1)}(\theta |h,{\cal C}) | {\cal O}\right]$ to obtain $\hat\theta$.
    \item[{\bf (d)}] $E_2$-step: Compute the expected values $W^{l}_{ijk}$ and $W^{ll'}_{ijk}$.
    \item[{\bf (e)}] $M_2$-step: Maximize $E\left[l^{(2)}(h|\theta,{\cal C}) | {\cal O}\right]$ to obtain $\hat h$.
    \item[{\bf (f)}] Repeat {\bf (b)}-{\bf (e)} until convergence.
\end{enumerate}

\noindent Upon convergence, we obtain $\hat\Theta=(\hat\theta^\top,h^\top)^\top$.

\subsection{Variance--covariance matrix estimation}

\noindent
The variance--covariance matrix of the joint estimator
$
\hat{\Theta} = (\hat{\theta}^{\top}, \hat{h}^{\top})^{\top}
$
is obtained using standard asymptotic theory for M-estimators. In the proposed
estimation procedure, the marginal parameters and the dependence parameters are
estimated through two coupled estimating equations. Let
\[
M_I(\Theta)=\frac{1}{I}\sum_{i=1}^{I} m_i(\Theta)
\]
denote the empirical estimating function, where \(m_i(\Theta)\) is the contribution of family \(i\). The estimator
\(\hat{\Theta}\) is defined as the solution of
\[
M_I(\hat{\Theta})=0.
\]

\noindent
For the competing-risk model considered here, the stacked estimating function is
defined as
\[
m_i(\Theta)
=
\begin{pmatrix}
\dfrac{\partial \ell_i^{(1)}(\theta \mid h, O_i)}
{\partial \theta}
\\[1.2em]
\dfrac{\partial \ell_i^{(2)}(h \mid \theta, O_i)}
{\partial h}
\end{pmatrix}.
\]
Here, \(O_i\) represents all observed data for family \(i\), including event times,
event types, censoring indicators, covariates, and kinship information. Under standard regularity conditions, a first-order Taylor expansion of
\(M_I(\hat{\Theta})\) around the true parameter value \(\Theta_0\) gives
\[
\sqrt{I}(\hat{\Theta}-\Theta_0)
=
A^{-1}
\left\{
\frac{1}{\sqrt{I}}\sum_{i=1}^{I}m_i(\Theta_0)
\right\}
+o_p(1),
\]
where
\[
A
=
\mathbb{E}
\left[
-\frac{\partial M_I(\Theta)}
{\partial \Theta^{\top}}
\right]_{\Theta=\Theta_0}.
\]

\noindent
By the central limit theorem,
\[
\frac{1}{\sqrt{I}}\sum_{i=1}^{I}m_i(\Theta_0)
\xrightarrow{\mathcal{L}}
\mathcal{N}(0,B),
\]
with
\[
B
=
\mathbb{E}
\left[
M_I(\Theta_0)M_I(\Theta_0)^{\top}
\right].
\]
It follows that
\[
\sqrt{I}(\hat{\Theta}-\Theta_0)
\xrightarrow{\mathcal{L}}
\mathcal{N}
\left(
0,
A^{-1}B(A^{-1})^{\top}
\right).
\]

\noindent
In practice, \(A\) and \(B\) are replaced by their empirical counterparts evaluated
at \(\hat{\Theta}\):
\[
\hat{A}
=
-\frac{1}{I}
\sum_{i=1}^{I}
\left.
\frac{\partial m_i(\Theta)}
{\partial \Theta^{\top}}
\right|_{\Theta=\hat{\Theta}},
\qquad
\hat{B}
=
\frac{1}{I}
\sum_{i=1}^{I}
m_i(\hat{\Theta})m_i(\hat{\Theta})^{\top}.
\]

\noindent
The sandwich estimator of the covariance matrix of \(\hat{\Theta}\) is
therefore given by
\[
v(\hat{\Theta})
=
\frac{1}{I}
\hat{A}^{-1}
\hat{B}
(\hat{A}^{-1})^{\top}.
\]

%% file: Resultats.tex
%\section{Some illustrations with the parameters of the marginal model}
%Brca1(Brca=1)
\section{Application to BRCA1 families from BCFR}

\subsection{Analysis}
\noindent A total of 1,927 individuals from 373 BRCA1 familiy with last follow-up ages ranging from 18.14 to 102.5 years are included in the dataset. The number of individuals per family is provided in Appendix~J. A family is included in the datset if and only if the proband had been diagnosed with either breast or ovarian cancer prior to or at the study entry. In total, 706 breast cancers (minimum age: 21.84 years; mean age: 44.09 years; maximum age: 86.00 years) and 132 ovarian cancers  (minimum age: 29.00 years; mean age: 54.11 years; maximum age: 89.00 years) were observed as first events. Finally, 837 of the 1,927 women were BRCA1 mutation carriers. 

\noindent We applied the  doubly-EM-algorithm to the data at hand to estimate the parameters of the proposed competing risks copula model. In this application, $T_{ij}$ is calculated starting at age 18, since no events are observed before that age. The estimated parameters and their variances are reported in Table~1. 

\begin{table}[H]
\centering
\footnotesize
\begin{tabular}{|c|c|c|}
\hline
Parameters & Estimate & se \\
\hline

$\log(\alpha_1)$ 
& 1.070 & 0.0387 \\

$\log(\lambda_1)$ 
& -13.278 & 0.5738 \\

$\beta_{1}$ 
& 1.830 & 0.1966 \\

$\log(\alpha_2)$ 
& 1.093 & 0.0629\\

$\log(\lambda_2)$ 
& -14.392 & 0.9376 \\

$\beta_{2}$ 
& 0.593 & 0.3043 \\

$h_{11}$ 
& 0.378 & 0.1198 \\

$h_{12}$ 
& $2.02\times10^{-5}$ & $1.74\times10^{-5}$ \\

$h_{22}$ 
& 0.564 & 0.5281 \\

\hline
\end{tabular}

\caption{Parameter estimates and variances for the proposed competing risks copula model. The competing risks correspond to breast  ($\varepsilon=1$) and ovarian cancers ($\varepsilon=2$) as first events.}
\label{tab:param_estimates}
\end{table}

\noindent Positive gene carrier status effects were observed for both cancers, with a substantially stronger magnitude for breast  ($\hat\beta_{1}=1.830$) than for ovarian cancer ($\hat\beta_{2}=0.593$). From these results, we computed the expected mean age at onset of the first cancer for both carriers and non-carriers and found $\hat E(T_{ij}|X_{ij}=0) \approx 93$ and $\hat E(T_{ij}|X_{ij}=1) \approx 61$. This highlights the huge effect of the gene carrier status. Note that these model-based estimates are higher than the observed mean ages at onset. This was expected since the empirical quantities are estimates of $E(T_{ij}|T_{ij}<C_{ij}, X_{ij}=0)$ and $E(T_{ij}|T_{ij}<C_{ij}, X_{ij}=1)$, respectively. 

\noindent As shown in Table~2, we also computed the estimates of the cause-specific expected mean ages $E(T_{ij}|\varepsilon_{ij}=l,X_{ij}=x)$ and found
%\begin{center}
\begin{table}[H]
\centering
\begin{tabular}{ |c|c|c| } 
 \hline
  & Breast cancer & Ovarian cancer \\ 
  \hline
 Non carriers $X=0$ & 92.53 & 93.31 \\ 
 Carriers $X=1$ & 61.43 & 61.88 \\ 
 \hline
\end{tabular}
\caption{Expected mean ages at onset. The competing risks correspond to breast $(\varepsilon = 1)$ and ovarian cancers $(\varepsilon = 2)$ as first events. Estimates are reported with respect to BRCA1 carrier status.}
\end{table}
%\end{center}

\noindent From these results, one notices that the estimates of $E(T_{ij}|\varepsilon_{ij}=l,X_{ij}=x)$ do not vary with $l$. This suggests that the distributions of the ages at onset of cancer as first event are the same for the breast and the ovarian cancers. Thereafter, we computed the estimates of $P(\varepsilon_{ij}=l|X_{ij}=x_{ij})$. The corresponding results are reported in Table~3.

%\begin{center}
\begin{table}[H]
\centering
\begin{tabular}{ |c|c|c| } 
 \hline
  & Breast cancer & Ovarian cancer \\ 
  \hline
 Non carriers $X=0$ & 0.691 & 0.309 \\ 
 Carriers $X=1$ & 0.889 & 0.111 \\ 
 \hline
\end{tabular}
\caption{Estimated probabilities of each cancer type as first event. The competing risks correspond to breast $(\varepsilon = 1)$ and ovarian cancers $(\varepsilon = 2)$ as first events. Estimates are reported with respect to BRCA1 carrier status.}
\end{table}
%\end{center}

\noindent Based on these results, we conclude that the risk of breast cancer to occur as the first cancer for this population is much higher than that of ovarian cancer, and this is even more pronounced for the \textit{BRCA1} carriers.

\noindent The estimated Weibull shape parameters exceed one for both event types, indicating increasing age-related hazards. In Figures~1 and~2, we present the estimated cumulative incidence functions for breast and ovarian cancer as first cancer events among BRCA1 mutation carriers and non-carriers, respectively, together with their 95\% confidence bands obtained using the delta method.

\begin{figure}[H]
    \centering 
        \includegraphics[scale=0.85]{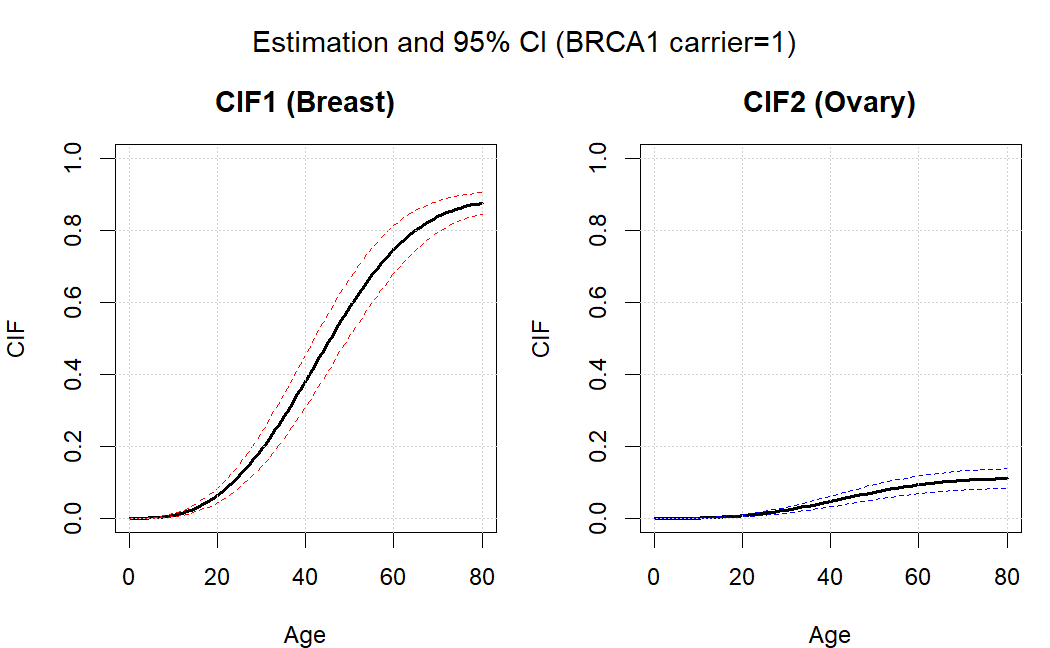}
         \caption{Estimated cumulative incidence functions among BRCA1 carriers with 95\% confidence intervals}
         \label{hiera}
    \label{fig:my_label}
\end{figure}

\begin{figure}[H]
    \centering 
        \includegraphics[scale=0.85]{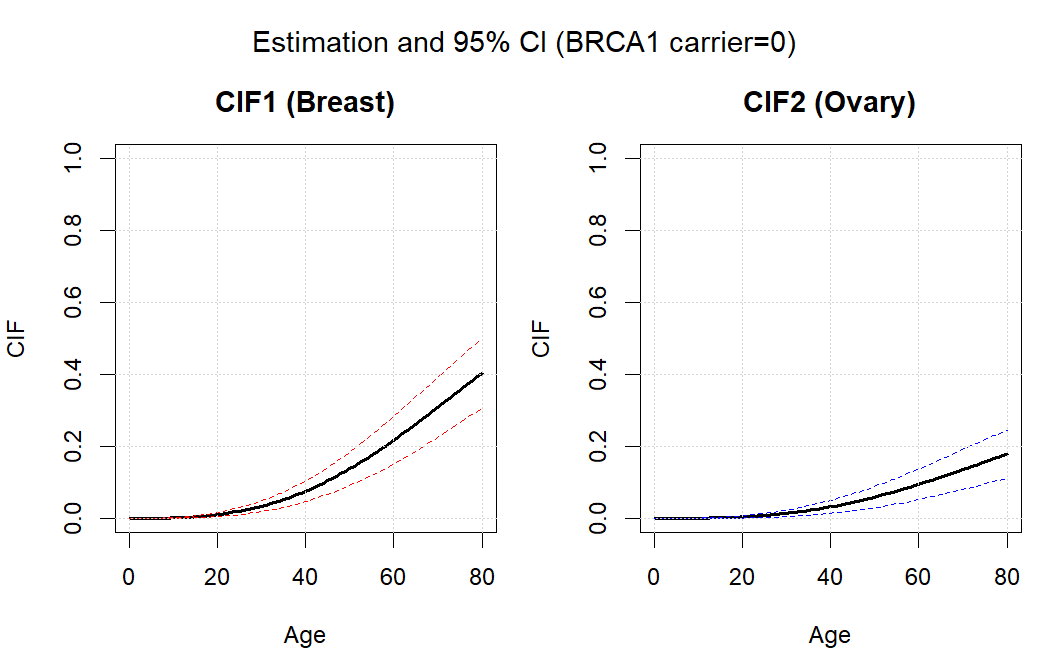}
         \caption{Estimated cumulative incidence functions among BRCA1 non-carriers with 95\% confidence intervals}
         \label{hiera}
    \label{fig:my_label}
\end{figure}

\noindent These figures show that the cumulative incidence of breast cancer as a first event is substantially higher among BRCA1 mutation carriers than among non-carriers across all ages. In contrast, the increase in ovarian cancer risk associated with BRCA1 mutations appears more moderate, although mutation carriers still exhibit consistently higher cumulative incidence functions over time.

\noindent Relatively moderate to strong familial dependences were observed for breast  ($h_{11}=0.378$) and ovarian cancers ($h_{22}=0.564$). In contrast,the cross-dependence parameter was nearly null ($h_{12}=2.02\times10^{-5}$), suggesting negligible residual dependence between the two first-event processes after adjustment for marginal effects and the familial structure. 

\noindent Finally, the variances estimation indicate satisfactory precision, although greater uncertainty was observed for $h_{22}$. This is due to the fact that the proportion of pairs of individuals from the same family who have both had ovarian cancer, i.e.  $(\delta_{ij}=\delta_{ik}=2)$, is very small.

\subsection{Practical illustrations}
\noindent To further investigate the temporal dynamics of the competing risks of breast and ovarian cancers, we consider several dependence measures and risk functions derived from the proposed model. These quantities assess marginal and dependence patterns over time. Specifically, we investigate the marginal risk ratio, the cross-odds ratio, the instantaneous cross-odds ratio, and the dynamic conditional risk function. The corresponding quantities are deduced from the estimates of the marginal and the dependence parameters of our model using the plug-in principle, where $\Theta=(\theta^\top,h^\top)^\top$ is replaced by $\hat{\Theta}=(\hat{\theta}^\top,\hat{h}^\top)^\top$. %Confidence bands for the estimated association measures are constructed by propagating $\mathrm{Var}(\hat{\Theta})$ through the corresponding functions using the delta method.

\subsubsection{Cause-specific cross ratio}

\noindent To characterize the temporal predominance of a specific cause $1 $, we define the \emph{marginal odds ratio} as
\begin{align*}
R(t,x)&=\frac{P(\varepsilon_{ij}=1 \mid T_{ij}=t,X_{ij}=x)}{P(\varepsilon_{ij}=1 \mid T_{ij} > t,X_{ij}=x)} 
\times \frac{P(\varepsilon_{ij}=2 \mid T_{ij} > t,X_{ij}=x)}{P(\varepsilon_{ij}=2 \mid T_{ij}=t,X_{ij}=x)},
\end{align*} 

\noindent The  expression of $R(t,x)$ in terms of the parameters $\Theta$ of our model is given in Appendix B.

\noindent The rationale for this dependence measure comes from Oakes (1989)\citep{oakes1989bivariate}. Indeed, one may see the ratio $P(\varepsilon_{ij}=1 \mid T_{ij}=t,X_{ij}=x)/P(\varepsilon_{ij}=1 \mid T_{ij}>t,X_{ij}=x)$ as a time-varying association measure between $T_{ij}$ and $\varepsilon_{ij}$. The Cause-specific cross ratio is the ratio of the measures for breast and ovarian cancers, respectively. 

\noindent A value of $R(t,x)$ greater than 1 indicates that women with covariates $x$ observe breast cancer as first event at younger ages than ovarian cancer. On the other hand, $R(t,x)=1$ indicates that both cancers are equally likely to occur as first event at age $t$. 
%\noindent This quantity compares the relative contribution of breast cancer versus ovarian cancer among events occurring exactly at age $t$ to the corresponding relative contribution among events that remain to occur after age $t$. Values greater than 1 indicate that breast cancer is relatively overrepresented among first cancer events occurring at age $t$, whereas values lower than 1 indicate a relative predominance of ovarian cancer among events occurring later in life.

\noindent Figure~3 displays the $R(t,x)$ as a function of age $t$, separately for $x=0$ (blue curves) and $x=1$ (red curves). 

\begin{figure}[H]
\centering 
\includegraphics[scale=0.65]{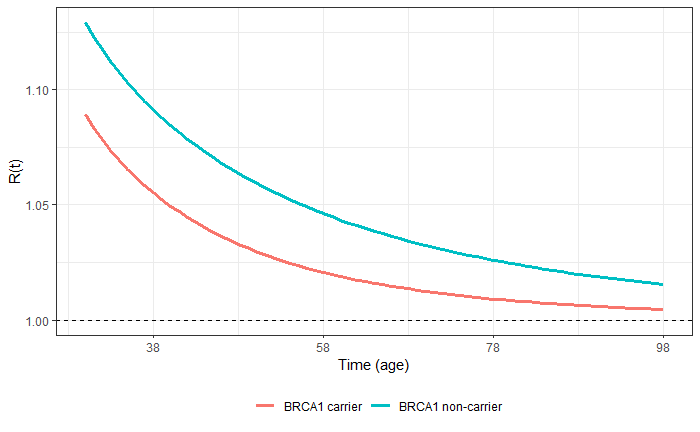}
\caption{Time-varying odds ratio comparing breast and ovarian cancer as first competing events.}
\end{figure}

\noindent The estimated curves remain above 1 over the entire age range, indicating that breast cancer is relatively more represented among first cancer events occurring at age $t$ than among subjects remaining event-free beyond age $t$. However, the temporal odds ratio decreases progressively toward 1 with increasing age, suggesting a gradual reduction in the temporal imbalance between breast and ovarian cancer as first events over time. The temporal odds ratio is consistently lower among \emph{BRCA1} mutation carriers, indicating that ovarian cancer becomes relatively competitive at earlier ages in carriers compared with non-carriers.

%%%%%%%%%%%%%%%%%%%%%%%%%%%%%%%%%%%%%%%%%%%%%%%%%%%%%%%%%%%%%%%%%%
\subsubsection{Cumulative cross-odds ratio}

\noindent Consider two individuals $j$ and $k$ from family $i$. Following Scheike and Sun (2012)\citep{scheike2012cross}, one may measure the dependences between the survival outcomes of individuals $j$ and $k$ from family $i$ using the \emph{cross-odds ratio} \citep{su2020association}
\begin{align*}
\pi_{l,l',x,x'}(t,s)
=\frac{\mathrm{ODDS}\big(T_{ij} \le t, \varepsilon_{ij} = l \mid T_{ik} \le s, \varepsilon_{ik} = l', X_{ij}=x, X_{ik}=x'\big)}{\mathrm{ODDS}\big(T_{ij} \le t, \varepsilon_{ij} = l \mid X_{ij}=x)},
\end{align*}
where, for any events $A$ and $B$, $\mathrm{ODDS}(A) = \frac{P(A)}{1 - P(A)}$ and $\mathrm{ODDS}(A \mid B) = \frac{P(A \mid B)}{1 - P(A \mid B)}.$  The expression of $\pi_{l,l',x,x'}(t,s)$ in terms of the parameters $\Theta$ of our model is given in Appendix C.

\noindent This quantity is a time-varying dependence measure between event types $l$ and $l'$ at times $t$ and $s$. A value equal to 1 corresponds to independence, while a value greater (resp. lower) than 1 indicates a positive (resp. negative) association.

\noindent In Figures 4 and 5, we present the estimated $\pi_{l,l',x,x'}(t,s)$ as a function of $t$ for $s\in\{30,35,40,45\}$, $x=x'=1$ (mutation carriers), $l=l'=1$ (Figure~4), and  $l=l'=2$ (Figure~5). Each figure displays the strength of the intrafamilial association as a function of the ages of the two individuals, for different values of the kinship coefficient (0.25, 0.125, 0.0625 and 0). Figures corresponding to the other values of $(l,l')$, together with some of their 3D surface representations, are provided in Appendix F.

\begin{figure}[H]
    \centering 
        \includegraphics[scale=0.8]{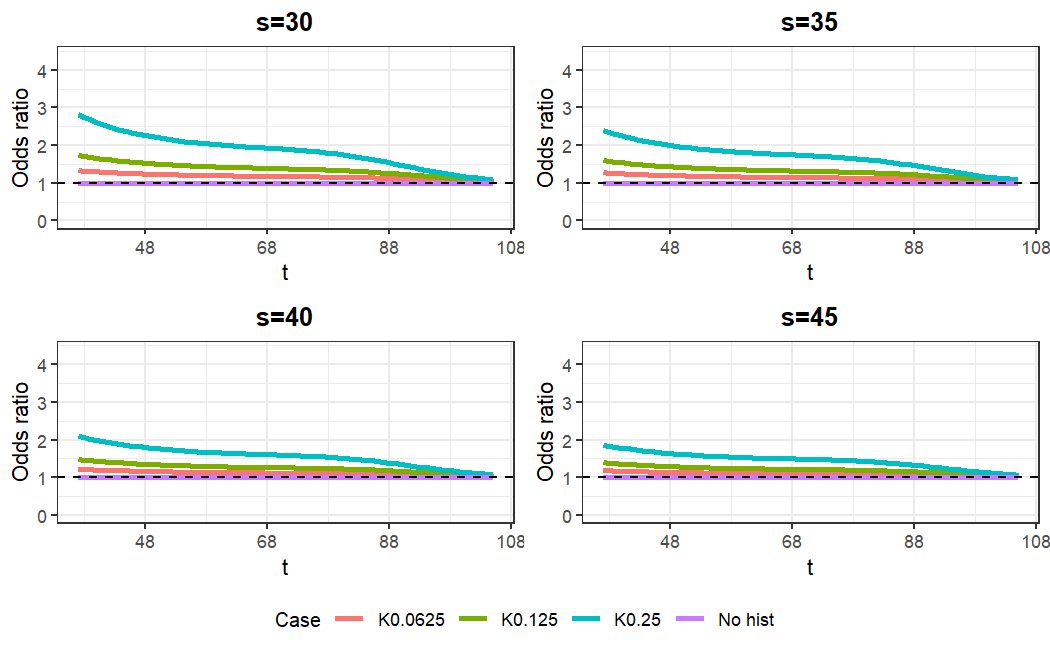}
        \caption{Estimated cumulative cross-odds ratio for breast cancer $(E_j=1, E_k=1)$ among BRCA1 carriers under different s.}
         \label{hiera}
    \label{fig:my_label}
\end{figure}

\begin{figure}[H]
    \centering 
        \includegraphics[scale=0.8]{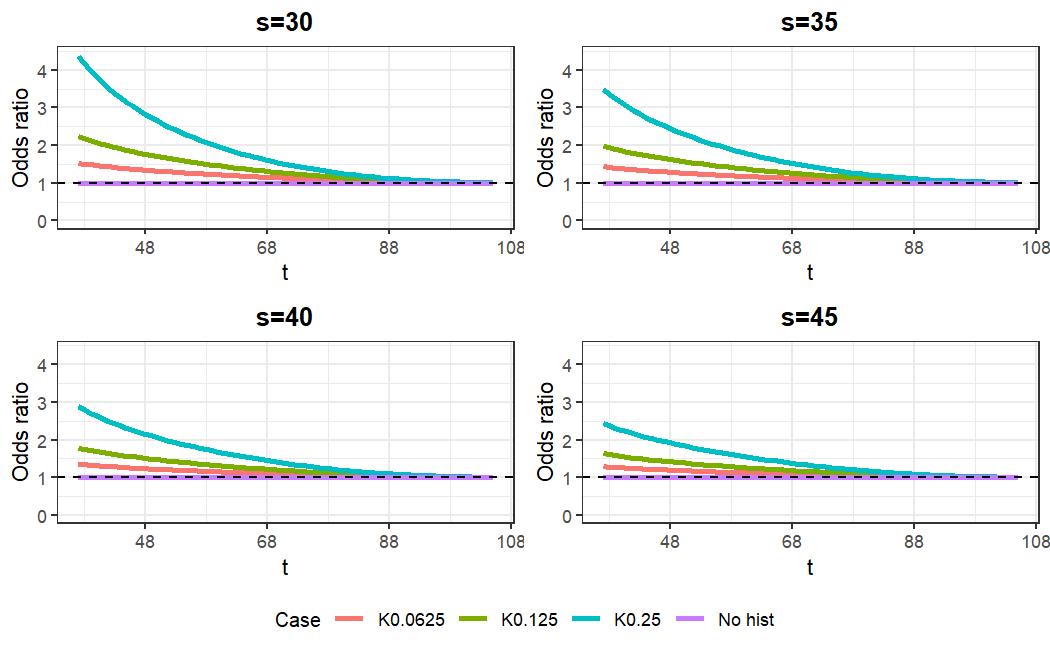}
        \caption{Estimated cumulative cross-odds ratio for ovarian cancer $(E_j=2, E_k=2)$ among BRCA1 carriers under different s.}
         \label{hiera}
    \label{fig:my_label}
\end{figure}

\noindent These figures illustrate the strong intrafamilial dependence highlighted by the values of $\pi_{l,l',x,x'}(t,s)$ among \emph{BRCA1} mutation carriers. For breast cancer as the first event (Figure~4) as well as for ovarian cancer (Figure~5), a marked positive association is observed, which increases with the kinship coefficient. This association reaches its highest values at a kinship of 0.25 (first-degree relatives), decreases progressively at lower kinship levels (0.125 and 0.0625), and disappears completely among unrelated individuals (kinship = 0).
These results demonstrate a substantial residual familial aggregation beyond the effect of the \emph{BRCA1} mutation itself, for both cancer types.

%%%%%%%%%%%%%%%%%%%%%%%%%%%%%%%%%%%%%%
\subsubsection{Instantaneous cross-odds ratio}

\noindent We define the \emph{instantaneous cross-odds ratio} as

\begin{align*}
\mathrm{OR}_{x,x'}(t,s)&=\frac{P_c(\varepsilon_{ij} = 1, \varepsilon_{ik} = 1)P_c(\varepsilon_{ij} = 2, \varepsilon_{ik} = 2)}{P_c(\varepsilon_{ij} = 1, \varepsilon_{ik} = 2 )P_c(\varepsilon_{ij} = 2, \varepsilon_{ik} = 1)
},
\end{align*} 
\noindent where $P_c(A)=P(A \mid T_{ij}=t,T_{ik}=s, X_{ij}=x,X_{ik}=x')$ for any event $A$. The explicit expression of $\mathrm{OR}_{x,x'}(t,s)$ in terms of the parameters $\Theta$ of our model is given in Appendix D.

\noindent Unlike the cumulative cross-odds ratio, this quantity measures the \emph{local} association between cancer types conditionally on the exact ages of occurrence $(t,s)$. A value of $\mathrm{OR}_{x,x'}(t,s)$ greater than 1 indicates a tendency toward concordance (same cancer type in both individuals), while a value lower than 1 suggests a predominance of discordant types (different cancers). In Figure~6, we present the estimated $\mathrm{OR}_{x,x'}(t,s)$ as function of $t$ for $s \in \{30,35,40,45\}$, $x=x'=1$, and $\phi^{(i)}_{kj} \in \{0,0.0625,0.125,0.25\}$. 3D surface representations of $\mathrm{OR}_{x,x'}(t,s)$ in terms of $t$ and $s$ are provided in Appendix G.
\begin{figure}[H]
    \centering 
        \includegraphics[scale=0.8]{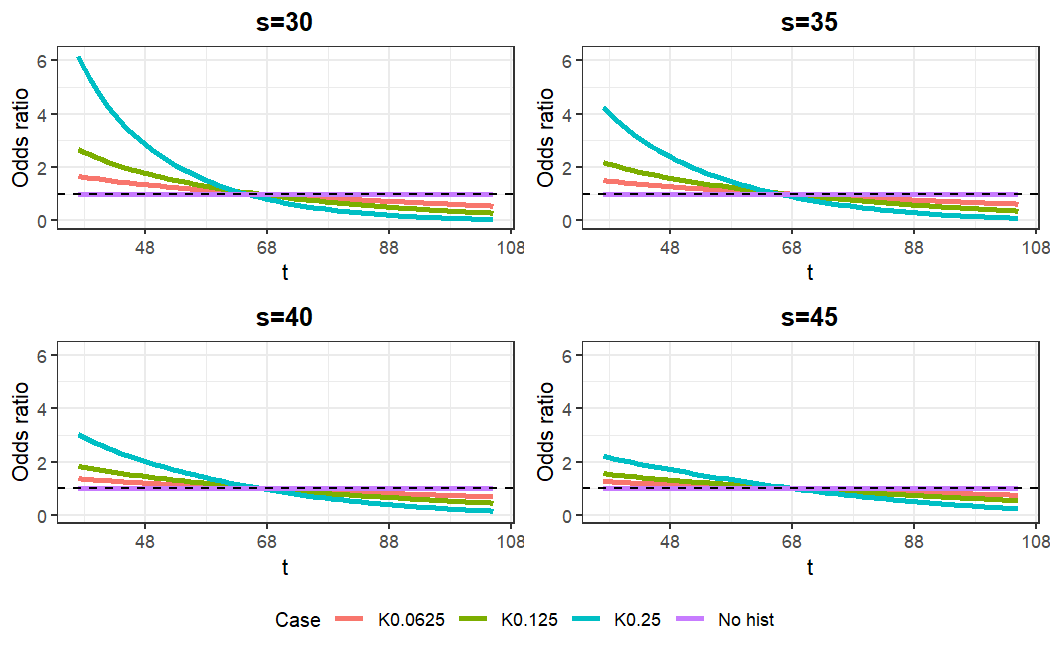}
 % \caption{Estimated instantaneous cross-odds ratio for identical cancer types among BRCA1 carriers}
  \caption{Instantaneous cross-odds ratio $\mathrm{OR}_{jk}(t,s)$ among \emph{BRCA1} mutation carriers, for different values of the kinship coefficient.}
         \label{hiera}
    \label{fig:my_label}
\end{figure}

\noindent
The results show that the association is strongest for the highest kinship value, \(\phi^{(i)}_{jk}=0.25\), particularly at younger ages and when \(t\) is close to the fixed value of \(s\). In this region, the odds ratio is above one, indicating a positive local association between the cancer types of two related individuals conditional on their exact event times.
As kinship decreases, the magnitude of the association is attenuated. The curves corresponding to \(\phi^{(i)}_{jk}=0.125\) and \(\phi^{(i)}_{jk}=0.0625\) remain closer to one, whereas the curve corresponding to no family history is essentially flat around one, as expected under independence. For larger values of \(t\), the odds ratio decreases and may fall below one, suggesting that the local concordance between event types weakens when the age at onset of individual \(j\) is substantially later than the fixed age \(s\) of individual \(k\).
These findings complement the cross-odds ratio by highlighting the local dependence structure conditional on the exact timing of the two events. While the cumulative measure summarizes dependence over time intervals, the instantaneous cross-odds ratio captures how the familial association varies locally with the event ages \(t\) and \(s\).

\subsubsection{Conditional dynamic risk}

\noindent We define the conditional dynamic risk as
$$CR_{s,t_0,l,l',x,x'}(t) = P(T_{ij} \le t, \varepsilon_{ij}=l \mid T_{ij}>s, T_{ik}=t_0, \varepsilon_{ik}=l',X_{ij}=x, X_{ik}=x')$$
for $0 \le s \le t$. \citep{su2020association}

\noindent It is equal to the conditional probability that individual \(j\) experiences an event of type l before age \(t\), given that the individual has remained event-free until age \(s\), and that the individual \(k\) experienced the event of type \(l'\) at the exact age \(t_0\). This quantity provides a dynamic measure of individualized risk accounting simultaneously for competing risks and familial dependence through the copula structure. The explicit expression of $CR_{s,t_0,l,l',x,x'}(t)$ in terms of the parameters $\Theta$ of our model is given in Appendix E.

\noindent The figure~7 shows $CR_{s,t_0,l,l',x,x'}(t)$ as a function of $t$, for $s=30$, $t_0=45$, $x=x'=1$, $l=2$, and $\phi^{(i)}_{kj} \in \{0,0.0625,0.125,0.25\}$. The left and right-hand panels correspond to $l'=1$ and $l'=2$, respectively.  Figures corresponding to $l=1$ are given in the Appendix H.

\begin{figure}[H]
    \centering 
        \includegraphics[scale=0.9]{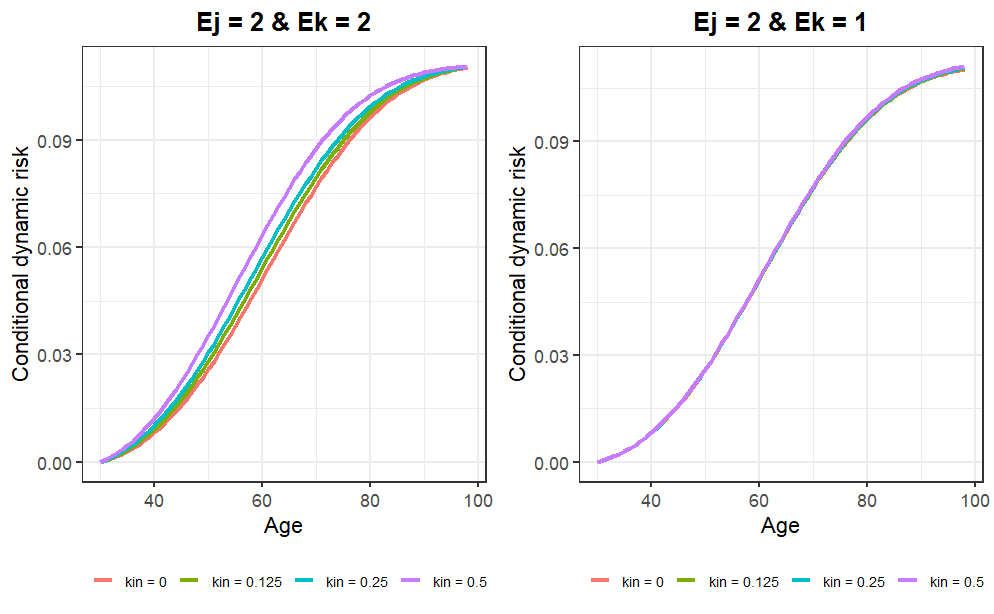}
         \caption{Estimated conditional dynamic risk of ovarian cancer for different kinship coefficients ($0.25$, $0.125$, $0.0625$, and $0$) for BRCA1. carrier}
         \label{hiera}
    \label{fig:my_label}
\end{figure}

\noindent The conditional dynamic risk of ovarian cancer given the relative has also ovarian cancer among \textit{BRCA1} carriers increases markedly with the degree of kinship to an affected relative. For first-degree relatives ($\phi^{(i)}_{kj} = 0.25$), the risk is substantially more elevated as compared to more distant relatives, and converges toward the \textit{BRCA1}-specific baseline when $\phi^{(i)}_{kj}=0$.

\noindent On the other hand, The conditional dynamic risk of ovarian cancer given the relative has had breast cancer don't vary with the kinship coefficient $\phi_{jk}^{(i)}$ among \textit{BRCA1} carriers. This was expected since we have obtained $\hat h_{12} \approx 0$.

\noindent A differential effect according to the relative's cancer type is observed. Having a relative with ovarian cancer (left-hand panel) consistently increases ovarian cancer risk (left-hand panel). This pattern indicates a stronger site-specific familial aggregation for ovarian cancer, even within the high-risk \textit{BRCA1} population.

\noindent The same pattern is observed for breast cancer. However, the extent is less pronounced since $\hat h_{11}$ is smaller than $\hat h_{22}$.

%For breast cancer risk (top panel, \textbf{why top panel?}), the pattern is more nuanced: the risk appears slightly higher when the relative had breast cancer (red curve) at younger ages among close relatives, although the two curves tend to converge at older ages. These findings suggest the presence of residual familial effects beyond the \textit{BRCA1} mutation itself, possibly due to genetic modifiers or other shared variants.

\noindent Overall, the results demonstrate that, even among \textit{BRCA1} carriers, detailed family history information, including both the degree of kinship and the specific cancer type in the affected relative, provides additional discriminatory power for refined dynamic risk prediction.

%% file: Simulation.tex
\section{Simulations}

\noindent We conducted a simulation study to evaluate the finite-sample performance of the proposed estimation procedure in a familial competing-risks setting. We considered a setting with $I$ independent families having all an identical structure. Each family consisted of 6 individuals. Specifically, the family structure was composed of 3 pairs of siblings, and any two individuals who were not siblings were assumed to be cousins. Therefore, the kinship matrix is
\[
\phi=
\begin{pmatrix}
0.5 & 0.25 & 0.0625 & 0.0625 & 0.0625 & 0.0625 \\
0.25 & 0.5 & 0.0625 & 0.0625 & 0.0625 & 0.0625 \\
0.0625 & 0.0625 & 0.5 & 0.25 & 0.0625 & 0.0625 \\
0.0625 & 0.0625 & 0.25 & 0.5 & 0.0625 & 0.0625 \\
0.0625 & 0.0625 & 0.0625 & 0.0625 & 0.5 & 0.25 \\
0.0625 & 0.0625 & 0.0625 & 0.0625 & 0.25 & 0.5
\end{pmatrix}.
\]
%%%%%%%%%%%%%%%%%%%%%%%

\noindent We considered a set of two covariates $X_{ij}=(X_{ij1},X_{ij2})$. Here, $X_{ij1}$ is a binary variable such as $P(X_{ij1}=0)=P(X_{ij1}=1)=1/2$ and $X_{ij2}$ a uniform variable over $[0,4]$. Given these covariates, the hazard function of $T_{ij}$ is
$$\lambda(t|X_{ij})= \lambda_1(t|X_{ij}) + \lambda_2(t|X_{ij})$$
with $\lambda_l(t|X_{ij})=\lambda_{0l}(t) e^{\beta_{1l}X_{ij1} + \beta_{2l}X_{ij1})}$. We set $\lambda_0l(t)=\alpha_{l}\lambda_{l} t^{\alpha_l-1}$, $\lambda_1=\lambda_2=1/1000$, $\alpha_1=\alpha_2=4$, $\beta_{11}=\beta_{12}=1$ and $\beta_{21}=\beta_{22}=-1$.
\noindent The proband ages at entry $\{a_i, i=1,\cdots,I\}$ were generated from a Weibull distribution such that $E[a_i]=4$ and $Var[a_i]=2$, whereas the censoring times $\{C_{ij},i=1,\cdots,I, j=0\cdots,5\}$ were generated from a uniform distribution over $[5,10]$. Furthermore, we set the heritability parameters to $h_{11}=h_{22}=0.4$ and $h_{12}=h_{21}=0.2$.
%%%%%%%%%%%%%%%%%%%%%%%%%%%%%%%%%

\noindent Given the covariates, the age at entry, and the censoring times, we generated the survival outcomes with selection bias as follows.

\noindent For $i=1,\dots,I$:
\begin{enumerate}
%\item[$\bullet$] Generate the proband covariates $(X_{i11},X_{i12})$ and the non-proband covariates $(X_{ij1},X_{ij2})$, $j=2,\dots,6$.
\item[$\bullet$] Generate $\varepsilon_{i0}$ from a Bernoulli distribution with probability
\[
P(\varepsilon_{i0}=1\mid T_{i0}\le a_i, X_{i0})
=
\frac{F_1(a_i\mid X_{i0})}
{F_1(a_i\mid X_{i0})+F_2(a_i\mid X_{i0})}.
\]
\item[$\bullet$] Conditional on $\varepsilon_{i0}=l$, generate $v_{i0}$ from a uniform distribution over $[0,1]$ and solve
\[
P(T_{i0}>t\mid T_{i0}\le a_i, \varepsilon_{i0}=l,X_{i0})=1-\frac{F_l(t\mid X_{i11},X_{i12})}
{F_l(a_i\mid X_{i11},X_{i12})}=v_{i0}
\]
in $t$ to obtain $t_{i0}$.
%Equivalently,\[F_k(t_{i1}\mid X_{i11},X_{i12})=(1-v_{i1})F_k(a_i\mid X_{i11},X_{i12}).\]
%Because the cumulative incidence functions did not admit a closed-form inverse, $t_{i1}$ was obtained by numerical inversion.

\item[$\bullet$] Compute
\[
u_{i0}
=
P(T_{i0}>t_{i0}\mid \varepsilon_{i0},X_{i0})
=
1-\frac{F_{\varepsilon_{i0}}(t_{i0}\mid X_{i0})}
{F_{\varepsilon_{i0}}(\infty\mid X_{i0})},
\]
and 
\[
z_{i0}=\Phi^{-1}(u_{i0}).
\]
\item[$\bullet$] For $j=1,\dots,5$, generate $\varepsilon_{ij}$ from a Bernoulli distribution with probability
\[
P(\varepsilon_{ij}=1\mid X_{ij})=F_1(\infty\mid X_{ij}).
\]
\item[$\bullet$] Construct the covariance matrix $V_i$ using the vector of causes
\[
(\varepsilon_{i0},\dots,\varepsilon_{i5})
\]
and generate $(Z_{i1},\dots,Z_{i5})^\top$ from the conditional multivariate normal distribution given $Z_{i0}=z_{i0}$.

\item[$\bullet$] For $j=1,\dots,5$:
\begin{enumerate}
\item[{\it (i)}] Compute
\[
u_{ij}=\Phi(Z_{ij}).
\]

\item[{\it (ii)}] Conditional on $\varepsilon_{ij}=k$, solve
\[
P(T_{ij}>t\mid \varepsilon_{ij}=k,X_{ij})=u_{ij}
\]
in $t$ to obtain
\[
1-\frac{F_k(t_{ij}\mid X_{ij1},X_{ij2})}
{F_k(\infty\mid X_{ij1},X_{ij2})}
=
u_{ij}.
\]
%Equivalently,\[F_k(t_{ij}\mid X_{ij1},X_{ij2})=(1-u_{ij})F_k(\infty\mid X_{ij1},X_{ij2}).\]
%As for the proband, $t_{ij}$ was obtained by numerical inversion.

\item[{\it (iii)}] Compute
\[
Y_{ij}=\min(t_{ij},C_{ij}),
\qquad
\delta_{ij}=\mathbf{1}(t_{ij}<C_{ij})\times \varepsilon_{ij}.
\]
\end{enumerate}

\item[$\bullet$] For the proband, define similarly
\[
Y_{i0}=\min(t_{i0},a_i),
\qquad
\delta_{i0}=\mathbf{1}(t_{i0}<a_i)\times \varepsilon_{i0}.
\]
\end{enumerate}
More details of the simulations are provided in Appendix K. We generated 200 independent datasets following the algorithm described above. For each dataset, we estimated the paramters $\Theta$ of our model using the doubly-EM-algorithm. The variances of the estimators were obtained following the method detailed in Section 4.4. The results for $I=400$ and $I=800$ are reported in Table 4.
%\noindent We repeated this procedure to generate the simulation datasets. For each simulated dataset, we estimated the marginal parameters governing the two competing causes as well as the dependence parameters $(h_{11},h_{12},h_{22})$. Variance estimation was based on asymptotic $M$-estimation theory using a sandwich estimator. The empirical bias, empirical standard deviation, average estimated standard error, and confidence interval coverage were then evaluated across replications.

\begin{table}[H]
\centering
\small
\renewcommand{\arraystretch}{1.1}
\setlength{\tabcolsep}{4pt}
\resizebox{1\textwidth}{!}{%
\begin{tabular}{|c|c|cccc|cccc|}
\hline
Parameters & True value & \multicolumn{4}{c|}{I=400} & \multicolumn{4}{c|}{I=800} \\
\hline
 &  & Bias & Empirical var. & Asymptotic var. & Coverage & Bias & Empirical var. & Asymptotic var. & Coverage \\
\hline
$\log(\alpha_1)$     & 1.3863  & -0.0030 & 0.0005 & 0.0005 & 93.5\% & 0.0027 & 0.0002 & 0.0002 & 94.0\% \\
$\log(\lambda_1)$    & -6.9078 & 0.0202  & 0.0240 & 0.0224 & 93.0\% & -0.0159 & 0.0105 & 0.0112 & 94.5\% \\
$\beta_{11}$ (X1)    & 1.0000  & -0.0041 & 0.0015 & 0.0013 & 92.5\% & 0.0016  & 0.0007 & 0.0007 & 95.5\% \\
$\beta_{12}$ (X2)    & -1.0000 & -0.0001 & 0.0029 & 0.0032 & 96.5\% & -0.0025 & 0.0017 & 0.0016 & 93.5\% \\
$\log(\alpha_2)$     & 1.3863  & 0.0008  & 0.0005 & 0.0005 & 93.5\% & 0.0019  & 0.0002 & 0.0002 & 96.0\% \\
$\log(\lambda_2)$    & -6.9078 & 0.0041  & 0.0247 & 0.0223 & 93.0\% & -0.0133 & 0.0111 & 0.0112 & 95.0\% \\
$\beta_{21}$ (X1)    & 1.0000  & -0.0007 & 0.0014 & 0.0013 & 94.5\% & 0.0002  & 0.0007 & 0.0007 & 93.0\% \\
$\beta_{22}$ (X2)    & -1.0000 & -0.0063 & 0.0035 & 0.0032 & 91.0\% & -0.0018 & 0.0018 & 0.0016 & 93.0\% \\
$h_{11}$             & 0.4000  & -0.0035 & 0.0115 & 0.0095 & 89.0\% & -0.0005 & 0.0054 & 0.0046 & 93.0\% \\
$h_{12}$             & 0.2000  & -0.0075 & 0.0066 & 0.0051 & 90.0\% & 0.0036  & 0.0030 & 0.0026 & 95.0\% \\
$h_{22}$             & 0.4000  & -0.0006 & 0.0105 & 0.0093 & 93.5\% & 0.0002  & 0.0052 & 0.0047 & 91.5\% \\
\hline
\end{tabular}%
}
\caption{Simulation results for $I=400$ and $I=800$ families. Empirical bias, empirical variance, asymptotic variance and 95\% confidence interval coverage rates.}
\label{tab:simulation}
\end{table}

\noindent From Table 4, we observe that the proposed estimation procedure performs very well in finite samples. The parameters $\Theta$ of our model are consitently estimated in the presence of an ascertainment bias induced by the proband.

\noindent The sandwich estimator based on M-estimation theory provides reliable variance estimates, as the empirical variances and asymptotic variances are very close. Consequently, the empirical coverage rates of the 95\% confidence intervals are generally close to the nominal level, particularly for $I=800$ families.

\noindent As expected, increasing the number of families from $I=400$ to $I=800$ leads to a substantial reduction in variance and improved coverage rates. The dependence parameters $h_{11}$, $h_{12}$ and $h_{22}$ show slightly larger variability and marginally lower coverage at $I=400$, which is consistent with the fact that these parameters represent variance components and require larger sample sizes to achieve stable estimation.

%% file: Conclusion.tex
\section{Conclusion}

\noindent In this paper, we propose a model for clustered competing risks using copulas. The cause specific hazard functions are assumed to follow a Cox proportional hazards model whereas the joint distribution of the cause-specific failure times for related subjects is modelled using a Gaussian copula whose correlation matrix is expressed in terms of the kinship coefficients. Additionally, we develop a doubly-\textrm{EM} algorithm to estimate the parameters of our model in the presence of a selection bias. Simulations suggest that the iterative estimation algorithm works well and the variances are well estimated. Moreover, the proposed model allows one to express and estimate a rich variaty of dependence measures and risk functions \citep{cheng2008nonparametric,cheng2010association,cheng2012cumulative,scheike2012cross}.%[\textcolor{red}{REF : citer quelques references parmi les references 6 a 20 du papier de statistics in medicine 2020}].

\noindent In the proposed model, we made two strong assumptions. First, we assumed that the event indicators $\{\varepsilon_{i0}, \cdots, \varepsilon_{in_i}\}$ are independent given the covariates. This assumption can easily be relaxed as one may assume that $\{\varepsilon_{i0}, \cdots, \varepsilon_{in_i}\}$ follows a copula indexed with dependence parameters \citep{su2020association}. Furthermore, we specified a Weibull distribution for the baseline cause-specific hazard. The easiest way to relax this assumtion is to consider a piecewise constant baseline hazard function.

\noindent This work focused on the onset of breast or ovarian cancer as a first event. However, it is still possible to observe a second and even a third cancer after the first event. Extending the proposed model to express and estimate the dependence measures and the risk functions beyond the first event is subject of ongoing research.

\noindent This present work is limited to one binary covariate. Extending the doubly \textrm{EM}-algorithm to situations with complex covariates such as functional or time-varying covariates is challenging, as a regularization/penalization additional step may be required at each iteration. The practical implementation of such an estimating procedure is subject of ongoing research. 

\noindent The main methodological contribution of this work is the development of a doubly \textrm{EM}-algorithm to estimate the parameters of the proposed model. This algorithm can easily be adapted to other situations where one has an iterative estimation algorithm with missing data.

%% file: AnnexeV.tex
\section{Appendix}
\subsection{Univariate likelihood}
\subsubsection{Contribution of probands}
All probands have observed at least one event. %on distingue 4 contributions des probands.
\begin{enumerate}
    \item $t_{i0}<a_i$, $\delta_{i0}=1$
    \begin{align}
        &P(T_{i0}=t_{i0}, \varepsilon_{i0}=1|T_{i0}<a_i) \nonumber \\
        &=\dfrac{P(T_{i0}=t_{i0}, \varepsilon_{i0}=1,T_{i0}<a_i)}{P(T_{i0}<a_i)}\nonumber \\
        &=\dfrac{P(T_{i0}=t_{i0}, \varepsilon_{i0}=1)}{P(T_{i0}<a_i)}\nonumber \\
        &=\dfrac{f_1(t_{i0})}{1-S_T(a_i)}
    \end{align}
        \item $t_{i0}<a_i$, $\delta_{i0}=2$
    \begin{align}
        &P(T_{i0}=t_{i0}, \varepsilon_{i0}=2|T_{i0}<a_i) \nonumber \\
        &=\dfrac{P(T_{i0}=t_{i0}, \varepsilon_{i0}=2,T_{i0}<a_i)}{P(T_{i0}<a_i)}\nonumber \\
        &=\dfrac{P(T_{i0}=t_{i0}, \varepsilon_{i0}=2)}{P(T_{i0}<a_i)}\nonumber \\
        &=\dfrac{f_2(t_{i0})}{1-S_T(a_i)}
    \end{align}

    \item $t_{i0}=a_i$, $\delta_{i0}=1$
    \begin{align}
        P(T_{i0}=t_{i0}, \varepsilon_{i0}=1)=f_1(t_{i0})
    \end{align}
    \item $t_{i0}=a_i$, $\delta_{i0}=2$
    \begin{align}
        P(T_{i0}=t_{i0}, \varepsilon_{i0}=2)=f_2(t_{i0})
    \end{align}
\end{enumerate}

\subsubsection{Contribution of non-probands}
\noindent When at least one of the two events has been observed (i.e., \(\delta_{ij}=1\)), we distinguish 4 contributions from the non-probands.
\begin{enumerate}
    \item $\delta_{ij}=1$, $\delta_{i0}=1$
    \begin{align}
        &P(T_{ij}=t_{ij}, \varepsilon_{ij}=1|T_{i0}=t_{i0}, \varepsilon_{i0}=1)\nonumber \\
        &=\dfrac{P(T_{ij}=t_{ij}, \varepsilon_{ij}=1,T_{i0}=t_{i0}, \varepsilon_{i0}=1)}{P(T_{i0}=t_{i0}, \varepsilon_{i0}=1)}\nonumber \\
        &=\dfrac{P(T_{ij}=t_{ij}, T_{i0}=t_{i0}|\varepsilon_{ij}=1, \varepsilon_{i0}=1)P(\varepsilon_{ij}=1, \varepsilon_{i0}=1)}{P(T_{i0}=t_{i0}, \varepsilon_{i0}=1)}\nonumber \\
        &=D_{\alpha_{11}}^{11}\{S_1(t_{ij}),S_1(t_{i0})\}f_1(t_{ij})
    \end{align}

       \item $\delta_{ij}=1$, $\delta_{i0}=2$
    \begin{align}
        &P(T_{ij}=t_{ij}, \varepsilon_{ij}=1|T_{i0}=t_{i0}, \varepsilon_{i0}=2)\nonumber \\
        &=\dfrac{P(T_{ij}=t_{ij}, \varepsilon_{ij}=1,T_{i0}=t_{i0}, \varepsilon_{i0}=2)}{P(T_{i0}=t_{i0}, \varepsilon_{i0}=2)}\nonumber \\
        &=\dfrac{P(T_{ij}=t_{ij}, T_{i0}=t_{i0}|\varepsilon_{ij}=1, \varepsilon_{i0}=2)P(\varepsilon_{ij}=1, \varepsilon_{i0}=2)}{P(T_{i0}=t_{i0}, \varepsilon_{i0}=2)}\nonumber \\
           &=D_{\alpha_{12}}^{11}\{S_1(t_{ij}),S_2(t_{i0})\}f_1(t_{ij})
    \end{align}

        \item $\delta_{ij}=2$, $\delta_{i0}=1$
    \begin{align}
        &P(T_{ij}=t_{ij}, \varepsilon_{ij}=2|T_{i0}=t_{i0}, \varepsilon_{i0}=1)\nonumber \\
        &=\dfrac{P(T_{ij}=t_{ij}, \varepsilon_{ij}=2,T_{i0}=t_{i0}, \varepsilon_{i0}=1)}{P(T_{i0}=t_{i0}, \varepsilon_{i0}=1)}\nonumber \\
        &=\dfrac{P(T_{ij}=t_{ij}, T_{i0}=t_{i0}|\varepsilon_{ij}=2, \varepsilon_{i0}=1)P(\varepsilon_{ij}=2, \varepsilon_{i0}=1)}{P(T_{i0}=t_{i0}, \varepsilon_{i0}=1)}\nonumber \\
            &=D_{\alpha_{21}}^{11}\{S_2(t_{ij}),S_1(t_{i0})\}f_2(t_{ij})
    \end{align}

       \item $\delta_{ij}=2$, $\delta_{i0}=2$
    \begin{align}
        &P(T_{ij}=t_{ij}, \varepsilon_{ij}=2|T_{i0}=t_{i0}, \varepsilon_{i0}=2)\nonumber \\
        &=\dfrac{P(T_{ij}=t_{ij}, \varepsilon_{ij}=2,T_{i0}=t_{i0}, \varepsilon_{i0}=2)}{P(T_{i0}=t_{i0}, \varepsilon_{i0}=2)}\nonumber \\
        &=\dfrac{P(T_{ij}=t_{ij}, T_{i0}=t_{i0}|\varepsilon_{ij}=2, \varepsilon_{i0}=2)P(\varepsilon_{ij}=2, \varepsilon_{i0}=2)}{P(T_{i0}=t_{i0}, \varepsilon_{i0}=2)}\nonumber \\
            &=D_{\alpha_{22}}^{11}\{S_2(t_{ij}),S_2(t_{i0})\}f_2(t_{ij})
    \end{align}
   \noindent When none of the two events has been observed (i.e., \(\delta_{ij}=0\)), we distinguish 2 contributions. Each contribution is weighted by the weight.  \par $W^{u}_{ll'}=P(\varepsilon_{ij}=l|T_{ij}>t_{ij}, T_{i0}=t_{i0}, \varepsilon_{i0}=l')$ with $l={1,2}$ and $l'={1,2}$.
   %%% 5 %%
   \item $\delta_{ij}=0$, $\delta_{i0}=1$
    $$P(T_{ij}>t_{ij}, \varepsilon_{ij}=1|T_{i0}=t_{i0}, \varepsilon_{i0}=1)^{W^{u}_{11}} \ *\  P(T_{ij}>t_{ij}, \varepsilon_{ij}=2|T_{i0}=t_{i0}, \varepsilon_{i0}=1)^{W^{u}_{21}}$$  with
    \begin{align}
           &P(T_{ij}>t_{ij}, \varepsilon_{ij}=1|T_{i0}=t_{i0}, \varepsilon_{i0}=1)\nonumber \\
        &=\dfrac{P(T_{ij}>t_{ij}, \varepsilon_{ij}=1,T_{i0}=t_{i0}, \varepsilon_{i0}=1)}{P(T_{i0}=t_{i0}, \varepsilon_{i0}=1)}\nonumber \\
        &=\dfrac{P(T_{ij}>t_{ij}, T_{i0}=t_{i0}|\varepsilon_{ij}=1, \varepsilon_{i0}=1)P(\varepsilon_{ij}=1, \varepsilon_{i0}=1)}{P(T_{i0}=t_{i0}, \varepsilon_{i0}=1)}\nonumber \\
          &=D_{\alpha_{11}}^{01}\{S_1(t_{ij}),S_1(t_{i0})\}F_1(\infty|X_{ij})\\
\text{and} \nonumber\\
        &P(T_{ij}>t_{ij}, \varepsilon_{ij}=2|T_{i0}=t_{i0}, \varepsilon_{i0}=1)\nonumber \\
        &=\dfrac{P(T_{ij}>t_{ij}, \varepsilon_{ij}=2,T_{i0}=t_{i0}, \varepsilon_{i0}=1)}{P(T_{i0}=t_{i0}, \varepsilon_{i0}=1)}\nonumber \\
        &=\dfrac{P(T_{ij}>t_{ij}, T_{i0}=t_{i0}|\varepsilon_{ij}=2, \varepsilon_{i0}=1)P(\varepsilon_{ij}=2, \varepsilon_{i0}=1)}{P(T_{i0}=t_{i0}, \varepsilon_{i0}=1)}\nonumber \\
       &=D_{\alpha_{21}}^{01}\{S_2(t_{ij}),S_1(t_{i0})\}F_2(\infty|X_{ij})
    \end{align}
    
  %%% 6 %%
       \item $\delta_{ij}=0$, $\delta_{i0}=2$
     $$P(T_{ij}>t_{ij}, \varepsilon_{ij}=1|T_{i0}=t_{i0}, \varepsilon_{i0}=2)^{W^{u}_{12}} \ *\   P(T_{ij}>t_{ij}, \varepsilon_{ij}=2|T_{i0}=t_{i0}, \varepsilon_{i0}=2)^{W^{u}_{22}}$$  with
    \begin{align}
        &P(T_{ij}>t_{ij}, \varepsilon_{ij}=1|T_{i0}=t_{i0}, \varepsilon_{i0}=2)\nonumber \\
        &=\dfrac{P(T_{ij}>t_{ij}, \varepsilon_{ij}=1,T_{i0}=t_{i0}, \varepsilon_{i0}=2)}{P(T_{i0}=t_{i0}, \varepsilon_{i0}=2)}\nonumber \\
        &=\dfrac{P(T_{ij}>t_{ij}, T_{i0}=t_{i0}|\varepsilon_{ij}=1, \varepsilon_{i0}=2)P(\varepsilon_{ij}=1, \varepsilon_{i0}=2)}{P(T_{i0}=t_{i0}, \varepsilon_{i0}=2)}\nonumber \\
        &=D_{\alpha_{12}}^{01}\{S_1(t_{ij}),S_2(t_{i0})\}F_1(\infty|X_{ij})\\
 \text{and}\nonumber\\
       &P(T_{ij}>t_{ij}, \varepsilon_{ij}=2|T_{i0}=t_{i0}, \varepsilon_{i0}=2)\nonumber \\
        &=\dfrac{P(T_{ij}>t_{ij}, \varepsilon_{ij}=2,T_{i0}=t_{i0}, \varepsilon_{i0}=2)}{P(T_{i0}=t_{i0}, \varepsilon_{i0}=2)}\nonumber \\
        &=\dfrac{P(T_{ij}>t_{ij}, T_{i0}=t_{i0}|\varepsilon_{ij}=2, \varepsilon_{i0}=2)P(\varepsilon_{ij}=2, \varepsilon_{i0}=2)}{P(T_{i0}=t_{i0}, \varepsilon_{i0}=2)}\nonumber \\
        &=D_{\alpha_{22}}^{01}\{S_2(t_{ij}),S_2(t_{i0})\}F_2(\infty|X_{ij})
    \end{align}
We express:
 \begin{align*}
       W^{u}_{ll'}&=P(\varepsilon_{ij}=l|T_{ij}>t_{ij}, T_{i0}=t_{i0}, \varepsilon_{i0}=l')\\
       &=\dfrac{P(T_{ij}>t_{ij}, T_{i0}=t_{i0}, \varepsilon_{ij}=l, \varepsilon_{i0}=l')}{P(T_{ij}>t_{ij}, T_{i0}=t_{i0}, \varepsilon_{i0}=l')}\\
       &=\dfrac{P(T_{ij}>t_{ij}, T_{i0}=t_{i0}|\varepsilon_{ij}=l, \varepsilon_{i0}=l')P(\varepsilon_{ij}=l, \varepsilon_{i0}=l')}{\sum_{r=1}^{2}P(T_{ij}>t_{ij}, T_{i0}=t_{i0}, \varepsilon_{ij}=r, \varepsilon_{i0}=l')}\\
       &=\dfrac{P(T_{ij}>t_{ij}, T_{i0}=t_{i0}|\varepsilon_{ij}=l, \varepsilon_{i0}=l')P(\varepsilon_{ij}=l, \varepsilon_{i0}=l')}{\sum_{r=1}^{2}P(T_{ij}>t_{ij}, T_{i0}=t_{i0}| \varepsilon_{ij}=r, \varepsilon_{i0}=l')P(\varepsilon_{ij}=r, \varepsilon_{i0}=l')}
   \end{align*}
   avec $l={1,2}$ et $l'={1,2}$.
\end{enumerate}

\subsection{Bivariate likelihood}
\noindent We consider only pairs of individuals. The contributions of the pairs are conditioned on the information of their probands.\\  
We distinguish 8 contributions when each subject in a pair has observed at least one event. 
\begin{enumerate}
   \item $\delta_{ij}=1$, $\delta_{ik}=1$, $\delta_{i0}=1$
   \begin{align}
    & P(T_{ij}=t_{ij}, T_{ik}=t_{ik}, \varepsilon_{ij}=1, \varepsilon_{ik}=1|T_{i0}=t_{i0}, \varepsilon_{i0}=1)\nonumber \\
    &=\dfrac{P(T_{ij}=t_{ij}, T_{ik}=t_{ik}, \varepsilon_{ij}=1, \varepsilon_{ik}=1,T_{i0}=t_{i0}, \varepsilon_{i0}=1)}{P(T_{i0}=t_{i0}, \varepsilon_{i0}=1)})\nonumber \\
    &=\dfrac{P(T_{ij}=t_{ij}, T_{ik}=t_{ik}, T_{i0}=t_{i0}| \varepsilon_{ij}=1, \varepsilon_{ik}=1, \varepsilon_{i0}=1)P(\varepsilon_{ij}=1, \varepsilon_{ik}=1,\varepsilon_{i0}=1)}{P(T_{i0}=t_{i0}, \varepsilon_{i0}=1)})\nonumber \\
    &=D_{\alpha_{11}}^{111}\{S_1(t_{ij}), S_1(t_{ik}), S_1(t_{i0})\}f_1(t_{ij})f_1(t_{ik})\nonumber \nonumber\\
    &\propto D_{\alpha_{11}}^{111}\{S_1(t_{ij}), S_1(t_{ik}), S_1(t_{i0})\}
   \end{align}

      \item $\delta_{ij}=1$, $\delta_{ik}=1$, $\delta_{i0}=2$
   \begin{align*}
      & P(T_{ij}=t_{ij}, T_{ik}=t_{ik}, \varepsilon_{ij}=1, \varepsilon_{ik}=1|T_{i0}=t_{i0}, \varepsilon_{i0}=2)\nonumber \\
    &=\dfrac{P(T_{ij}=t_{ij}, T_{ik}=t_{ik}, \varepsilon_{ij}=1, \varepsilon_{ik}=1,T_{i0}=t_{i0}, \varepsilon_{i0}=2)}{P(T_{i0}=t_{i0}, \varepsilon_{i0}=2)})\nonumber \\
    &=\dfrac{P(T_{ij}=t_{ij}, T_{ik}=t_{ik}, T_{i0}=t_{i0}| \varepsilon_{ij}=1, \varepsilon_{ik}=1, \varepsilon_{i0}=2)P(\varepsilon_{ij}=1, \varepsilon_{ik}=1,\varepsilon_{i0}=2)}{P(T_{i0}=t_{i0}, \varepsilon_{i0}=2)})\nonumber \\
    &=D_{\alpha_{12}}^{111}\{S_1(t_{ij}), S_1(t_{ik}), S_2(t_{i0})\}f_1(t_{ij})f_1(t_{ik})\nonumber\\
    &\propto D_{\alpha_{12}}^{111}\{S_1(t_{ij}), S_1(t_{ik}), S_2(t_{i0})\}
   \end{align*}

      \item $\delta_{ij}=2$,  $\delta_{ik}=1$, $\delta_{i0}=1$
   \begin{align*}
    & P(T_{ij}=t_{ij}, T_{ik}=t_{ik}, \varepsilon_{ij}=2, \varepsilon_{ik}=1|T_{i0}=t_{i0}, \varepsilon_{i0}=1)\nonumber \\
    &=\dfrac{P(T_{ij}=t_{ij}, T_{ik}=t_{ik}, \varepsilon_{ij}=2, \varepsilon_{ik}=1,T_{i0}=t_{i0}, \varepsilon_{i0}=1)}{P(T_{i0}=t_{i0}, \varepsilon_{i0}=1)})\nonumber \\
    &=\dfrac{P(T_{ij}=t_{ij}, T_{ik}=t_{ik}, T_{i0}=t_{i0}| \varepsilon_{ij}=2, \varepsilon_{ik}=1, \varepsilon_{i0}=1)P(\varepsilon_{ij}=2, \varepsilon_{ik}=1,\varepsilon_{i0}=1)}{P(T_{i0}=t_{i0}, \varepsilon_{i0}=1)})\nonumber \\
    &=D_{\alpha_{21}}^{111}\{S_2(t_{ij}), S_1(t_{ik}), S_1(t_{i0})\}f_2(t_{ij})f_1(t_{ik})\nonumber\\
    &\propto D_{\alpha_{21}}^{111}\{S_2(t_{ij}), S_1(t_{ik}), S_1(t_{i0})\}
   \end{align*}

      \item $\delta_{ij}=2$, $\delta_{ik}=1$, $\delta_{i0}=2$
   \begin{align*}
      & P(T_{ij}=t_{ij}, T_{ik}=t_{ik}, \varepsilon_{ij}=2, \varepsilon_{ik}=1|T_{i0}=t_{i0}, \varepsilon_{i0}=2)\nonumber \\
    &=\dfrac{P(T_{ij}=t_{ij}, T_{ik}=t_{ik}, \varepsilon_{ij}=2, \varepsilon_{ik}=1,T_{i0}=t_{i0}, \varepsilon_{i0}=2)}{P(T_{i0}=t_{i0}, \varepsilon_{i0}=1)})\nonumber \\
    &=\dfrac{P(T_{ij}=t_{ij}, T_{ik}=t_{ik}, T_{i0}=t_{i0}| \varepsilon_{ij}=2, \varepsilon_{ik}=1, \varepsilon_{i0}=1)P(\varepsilon_{ij}=2, \varepsilon_{ik}=1,\varepsilon_{i0}=2)}{P(T_{i0}=t_{i0}, \varepsilon_{i0}=2)})\nonumber \\
    &=D_{\alpha_{21}}^{111}\{S_2(t_{ij}), S_1(t_{ik}), S_2(t_{i0})\}f_2(t_{ij})f_1(t_{ik})\nonumber\\
    &\propto D_{\alpha_{21}}^{111}\{S_2(t_{ij}), S_1(t_{ik}), S_2(t_{i0})\}
   \end{align*}

 \item  $\delta_{ij}=1$, $\delta_{ik}=2$, $\delta_{i0}=1$
   \begin{align*}
      & P(T_{ij}=t_{ij}, T_{ik}=t_{ik}, \varepsilon_{ij}=1, \varepsilon_{ik}=2|T_{i0}=t_{i0}, \varepsilon_{i0}=1)\nonumber \\
    &=\dfrac{P(T_{ij}=t_{ij}, T_{ik}=t_{ik}, \varepsilon_{ij}=1, \varepsilon_{ik}=2,T_{i0}=t_{i0}, \varepsilon_{i0}=1)}{P(T_{i0}=t_{i0}, \varepsilon_{i0}=1)})\nonumber \\
    &=\dfrac{P(T_{ij}=t_{ij}, T_{ik}=t_{ik}, T_{i0}=t_{i0}| \varepsilon_{ij}=1, \varepsilon_{ik}=2, \varepsilon_{i0}=1)P(\varepsilon_{ij}=1, \varepsilon_{ik}=2,\varepsilon_{i0}=1)}{P(T_{i0}=t_{i0}, \varepsilon_{i0}=1)})\nonumber \\
    &=D_{\alpha_{21}}^{111}\{S_1(t_{ij}), S_2(t_{ik}), S_1(t_{i0})\}f_1(t_{ij})f_2(t_{ik})\nonumber\\
    &\propto D_{\alpha_{21}}^{111}\{S_1(t_{ij}), S_2(t_{ik}), S_1(t_{i0})\}
   \end{align*}

      \item $\delta_{ij}=1$, $\delta_{ik}=2$, $\delta_{i0}=2$
   \begin{align*}
    & P(T_{ij}=t_{ij}, T_{ik}=t_{ik}, \varepsilon_{ij}=1, \varepsilon_{ik}=2|T_{i0}=t_{i0}, \varepsilon_{i0}=2)\nonumber \\
    &=\dfrac{P(T_{ij}=t_{ij}, T_{ik}=t_{ik}, \varepsilon_{ij}=1, \varepsilon_{ik}=2,T_{i0}=t_{i0}, \varepsilon_{i0}=2)}{P(T_{i0}=t_{i0}, \varepsilon_{i0}=2)})\nonumber \\
    &=\dfrac{P(T_{ij}=t_{ij}, T_{ik}=t_{ik}, T_{i0}=t_{i0}| \varepsilon_{ij}=1, \varepsilon_{ik}=2, \varepsilon_{i0}=2)P(\varepsilon_{ij}=1, \varepsilon_{ik}=2,\varepsilon_{i0}=2)}{P(T_{i0}=t_{i0}, \varepsilon_{i0}=2)})\nonumber \\
    &=D_{\alpha_{21}}^{111}\{S_1(t_{ij}), S_2(t_{ik}), S_2(t_{i0})\}f_1(t_{ij})f_2(t_{ik})\nonumber\\
    &\propto D_{\alpha_{21}}^{111}\{S_1(t_{ij}), S_2(t_{ik}), S_2(t_{i0})\}
   \end{align*}

    \item $\delta_{ij}=2$,  $\delta_{ik}=2$, $\delta_{i0}=1$
   \begin{align*}
      & P(T_{ij}=t_{ij}, T_{ik}=t_{ik}, \varepsilon_{ij}=2, \varepsilon_{ik}=2|T_{i0}=t_{i0}, \varepsilon_{i0}=1)\nonumber \\
    &=\dfrac{P(T_{ij}=t_{ij}, T_{ik}=t_{ik}, \varepsilon_{ij}=2, \varepsilon_{ik}=2,T_{i0}=t_{i0}, \varepsilon_{i0}=1)}{P(T_{i0}=t_{i0}, \varepsilon_{i0}=1)})\nonumber \\
    &=\dfrac{P(T_{ij}=t_{ij}, T_{ik}=t_{ik}, T_{i0}=t_{i0}| \varepsilon_{ij}=2, \varepsilon_{ik}=2, \varepsilon_{i0}=1)P(\varepsilon_{ij}=2, \varepsilon_{ik}=2,\varepsilon_{i0}=1)}{P(T_{i0}=t_{i0}, \varepsilon_{i0}=1)})\nonumber \\
    &=D_{\alpha_{21}}^{111}\{S_2(t_{ij}), S_2(t_{ik}), S_1(t_{i0})\}f_2(t_{ij})f_2(t_{ik})\nonumber\\
    &\propto D_{\alpha_{21}}^{111}\{S_2(t_{ij}), S_2(t_{ik}), S_1(t_{i0})\}
   \end{align*}

      \item $\delta_{ij}=2$, $\delta_{ik}=2$, $\delta_{i0}=2$
   \begin{align*}
      & P(T_{ij}=t_{ij}, T_{ik}=t_{ik}, \varepsilon_{ij}=2, \varepsilon_{ik}=2|T_{i0}=t_{i0}, \varepsilon_{i0}=2)\nonumber \\
    &=\dfrac{P(T_{ij}=t_{ij}, T_{ik}=t_{ik}, \varepsilon_{ij}=2, \varepsilon_{ik}=2,T_{i0}=t_{i0}, \varepsilon_{i0}=2)}{P(T_{i0}=t_{i0}, \varepsilon_{i0}=2)})\nonumber \\
    &=\dfrac{P(T_{ij}=t_{ij}, T_{ik}=t_{ik}, T_{i0}=t_{i0}| \varepsilon_{ij}=2, \varepsilon_{ik}=2, \varepsilon_{i0}=2)P(\varepsilon_{ij}=2, \varepsilon_{ik}=2,\varepsilon_{i0}=2)}{P(T_{i0}=t_{i0}, \varepsilon_{i0}=2)})\nonumber \\
    &=D_{\alpha_{22}}^{111}\{S_2(t_{ij}), S_2(t_{ik}), S_2(t_{i0})\}f_2(t_{ij})f_2(t_{ik})\nonumber\\
    &\propto D_{\alpha_{22}}^{111}\{S_2(t_{ij}), S_2(t_{ik}), S_2(t_{i0})\}
   \end{align*}
   
  \noindent When, for each pair, one of the subjects has not observed at least one event, namely subject \( ij \), we distinguish 4 contributions that are weighted by
  $W^{b,0}_{l,l',l_0}=P(\varepsilon_{ij}=l| T_{ij}>t_{ij}, T_{ik}=t_{ik}, \varepsilon_{ik}=l', T_{i0}=t_{i0}, \varepsilon_{i0}=l_0)$\\
       %%% 9 %%
\item $\delta_{ij}=0$, $\delta_{ik}=1$, $\delta_{i0}=1$
\begin{align*}
  P(T_{ij}>t_{ij}, T_{ik}=t_{ik}, \varepsilon_{ij}=1, \varepsilon_{ik}=1| T_{i0}=t_{i0}, \varepsilon_{i0}=1)^{W^{1}_{1,1,1}}\ *\\ \ P(T_{ij}>t_{ij}, T_{ik}=t_{ik}, \varepsilon_{ij}=2, \varepsilon_{ik}=1| T_{i0}=t_{i0}, \varepsilon_{i0}=1)^{W^{1}_{2,1,1}}  
\end{align*} with
     \begin{align*}
    &P(T_{ij}>t_{ij}, T_{ik}=t_{ik}, \varepsilon_{ij}=1, \varepsilon_{ik}=1| T_{i0}=t_{i0}, \varepsilon_{i0}=1)\nonumber \\
    &=\dfrac{P(T_{ij}>t_{ij}, T_{ik}=t_{ik}, \varepsilon_{ij}=1, \varepsilon_{ik}=1,T_{i0}=t_{i0}, \varepsilon_{i0}=1)}{P(T_{i0}=t_{i0}, \varepsilon_{i0}=1)})\nonumber \\
    &=\dfrac{P(T_{ij}>t_{ij}, T_{ik}=t_{ik}, T_{i0}=t_{i0}| \varepsilon_{ij}=1, \varepsilon_{ik}=1, \varepsilon_{i0}=1)P(\varepsilon_{ij}=1, \varepsilon_{ik}=1,\varepsilon_{i0}=1)}{P(T_{i0}=t_{i0}, \varepsilon_{i0}=1)})\nonumber \\
    &=D_{\alpha_{11}}^{011}\{S_1(t_{ij}), S_1(t_{ik}), S_1(t_{i0})\}f_1(t_{ik})F_1(\infty|X_{ij})\nonumber\\
    & \propto D_{\alpha_{11}}^{011}\{S_1(t_{ij}), S_1(t_{ik}), S_1(t_{i0})\} \ \text{and}
\end{align*}
\begin{align*}
        &P(T_{ij}>t_{ij}, T_{ik}=t_{ik}, \varepsilon_{ij}=2, \varepsilon_{ik}=1| T_{i0}=t_{i0}, \varepsilon_{i0}=1)\nonumber \\
    &=\dfrac{P(T_{ij}>t_{ij}, T_{ik}=t_{ik}, \varepsilon_{ij}=2, \varepsilon_{ik}=1,T_{i0}=t_{i0}, \varepsilon_{i0}=1)}{P(T_{i0}=t_{i0}, \varepsilon_{i0}=1)})\nonumber \\
    &=\dfrac{P(T_{ij}>t_{ij}, T_{ik}=t_{ik}, T_{i0}=t_{i0}| \varepsilon_{ij}=2, \varepsilon_{ik}=1, \varepsilon_{i0}=1)P(\varepsilon_{ij}=2, \varepsilon_{ik}=1,\varepsilon_{i0}=1)}{P(T_{i0}=t_{i0}, \varepsilon_{i0}=1)})\nonumber \\
    &=D_{\alpha_{21}}^{011}\{S_2(t_{ij}), S_1(t_{ik}), S_1(t_{i0})\}f_1(t_{ik})F_2(\infty|X_{ij})\nonumber\\
    &\propto D_{\alpha_{21}}^{011}\{S_2(t_{ij}), S_1(t_{ik}), S_1(t_{i0})\}
    \end{align*}
    
 %%% 10 %%
        \item $\delta_{ij}=0$, $\delta_{ik}=1$, $\delta_{i0}=2$
    \begin{align*}
        P(T_{ij}>t_{ij}, T_{ik}=t_{ik}, \varepsilon_{ij}=1, \varepsilon_{ik}=1| T_{i0}=t_{i0}, \varepsilon_{i0}=2)^{W^{1}_{1,1,2}} +\\
        P(T_{ij}>t_{ij}, T_{ik}=t_{ik}, \varepsilon_{ij}=2, \varepsilon_{ik}=1| T_{i0}=t_{i0}, \varepsilon_{i0}=2)^{W^{2}_{2,1,2}}
    \end{align*} with
    \begin{align*}
        &P(T_{ij}>t_{ij}, T_{ik}=t_{ik}, \varepsilon_{ij}=1, \varepsilon_{ik}=1| T_{i0}=t_{i0}, \varepsilon_{i0}=2)\nonumber \\
    &=\dfrac{P(T_{ij}>t_{ij}, T_{ik}=t_{ik}, \varepsilon_{ij}=1, \varepsilon_{ik}=1,T_{i0}=t_{i0}, \varepsilon_{i0}=2)}{P(T_{i0}=t_{i0}, \varepsilon_{i0}=2)})\nonumber \\
    &=\dfrac{P(T_{ij}>t_{ij}, T_{ik}=t_{ik}, T_{i0}=t_{i0}| \varepsilon_{ij}=1, \varepsilon_{ik}=1, \varepsilon_{i0}=2)P(\varepsilon_{ij}=1, \varepsilon_{ik}=1,\varepsilon_{i0}=2)}{P(T_{i0}=t_{i0}, \varepsilon_{i0}=2)})\nonumber \\
    &=D_{\alpha_{12}}^{011}\{S_1(t_{ij}), S_1(t_{ik}), S_2(t_{i0})\}f_1(t_{ik})F_1(\infty|X_{ij})\nonumber\\
    &\propto D_{\alpha_{12}}^{011}\{S_1(t_{ij}), S_1(t_{ik}), S_2(t_{i0})\}
    \\
 \text{and}, \nonumber\\
    &P(T_{ij}>t_{ij}, T_{ik}=t_{ik}, \varepsilon_{ij}=2, \varepsilon_{ik}=1| T_{i0}=t_{i0}, \varepsilon_{i0}=2)\nonumber \\
    &=\dfrac{P(T_{ij}>t_{ij}, T_{ik}=t_{ik}, \varepsilon_{ij}=2, \varepsilon_{ik}=1,T_{i0}=t_{i0}, \varepsilon_{i0}=2)}{P(T_{i0}=t_{i0}, \varepsilon_{i0}=1)})\nonumber \\
    &=\dfrac{P(T_{ij}>t_{ij}, T_{ik}=t_{ik}, T_{i0}=t_{i0}| \varepsilon_{ij}=2, \varepsilon_{ik}=1, \varepsilon_{i0}=1)P(\varepsilon_{ij}=2, \varepsilon_{ik}=1,\varepsilon_{i0}=2)}{P(T_{i0}=t_{i0}, \varepsilon_{i0}=2)})\nonumber \\
    &=D_{\alpha_{12}}^{011}\{S_2(t_{ij}), S_1(t_{ik}), S_2(t_{i0})\}f_1(t_{ik})F_2(\infty|X_{ij})\nonumber\\
    & \propto D_{\alpha_{12}}^{011}\{S_2(t_{ij}), S_1(t_{ik}), S_2(t_{i0})\}.
    \end{align*}

    %%%%%%%%%%%%%%%%%%%%%%%%%%%%%%%%%%%%%%%%%%%%%%%%%%%%%%%%%%%
     %%% 13 %%
    \item $\delta_{ij}=0$, $\delta_{ik}=2$, $\delta_{i0}=1$
        
    \begin{align*}
     P(T_{ij}>t_{ij}, T_{ik}=t_{ik}, \varepsilon_{ij}=1, \varepsilon_{ik}=2| T_{i0}=t_{i0}, \varepsilon_{i0}=1)^{W^{1}_{1,2,1}}  \ *\\
     P(T_{ij}>t_{ij}, T_{ik}=t_{ik}, \varepsilon_{ij}=2, \varepsilon_{ik}=2| T_{i0}=t_{i0}, \varepsilon_{i0}=1)^{W^{2}_{2,2,1}}
    \end{align*}
    with
    \begin{align*}
    &P(T_{ij}>t_{ij}, T_{ik}=t_{ik}, \varepsilon_{ij}=1, \varepsilon_{ik}=2| T_{i0}=t_{i0}, \varepsilon_{i0}=1)\nonumber \\
    &=\dfrac{P(T_{ij}>t_{ij}, T_{ik}=t_{ik}, \varepsilon_{ij}=1, \varepsilon_{ik}=2,T_{i0}=t_{i0}, \varepsilon_{i0}=1)}{P(T_{i0}=t_{i0}, \varepsilon_{i0}=1)})\nonumber \\
    &=\dfrac{P(T_{ij}>t_{ij}, T_{ik}=t_{ik}, T_{i0}=t_{i0}| \varepsilon_{ij}=1, \varepsilon_{ik}=2, \varepsilon_{i0}=1)P(\varepsilon_{ij}=1, \varepsilon_{ik}=2,\varepsilon_{i0}=1)}{P(T_{i0}=t_{i0}, \varepsilon_{i0}=1)})\nonumber \\
    &=D_{\alpha_{21}}^{011}\{S_1(t_{ij}), S_2(t_{ik}), S_1(t_{i0})\}f_2(t_{ik})F_1(\infty|X_{ij})\nonumber\\
    &\propto D_{\alpha_{21}}^{011}\{S_1(t_{ij}), S_2(t_{ik}), S_1(t_{i0})\}
    \\
 & \text{and}\nonumber\\
        &P(T_{ij}>t_{ij}, T_{ik}=t_{ik}, \varepsilon_{ij}=2, \varepsilon_{ik}=2| T_{i0}=t_{i0}, \varepsilon_{i0}=1)\nonumber \\
    &=\dfrac{P(T_{ij}>t_{ij}, T_{ik}=t_{ik}, \varepsilon_{ij}=2, \varepsilon_{ik}=2,T_{i0}=t_{i0}, \varepsilon_{i0}=1)}{P(T_{i0}=t_{i0}, \varepsilon_{i0}=1)})\nonumber \\
    &=\dfrac{P(T_{ij}>t_{ij}, T_{ik}=t_{ik}, T_{i0}=t_{i0}| \varepsilon_{ij}=2, \varepsilon_{ik}=2, \varepsilon_{i0}=1)P(\varepsilon_{ij}=2, \varepsilon_{ik}=2,\varepsilon_{i0}=1)}{P(T_{i0}=t_{i0}, \varepsilon_{i0}=1)})\nonumber \\
    &=D_{\alpha_{21}}^{011}\{S_2(t_{ij}), S_2(t_{ik}), S_1(t_{i0})\}f_2(t_{ik})F_2(\infty|X_{ij})\nonumber\\
    &\propto D_{\alpha_{21}}^{011}\{S_2(t_{ij}), S_2(t_{ik}), S_1(t_{i0})\}
    \end{align*}
 %%% 14 %%
        \item $\delta_{ij}=0$, $\delta_{ik}=2$, $\delta_{i0}=2$
    \begin{align*}
     P(T_{ij}>t_{ij}, T_{ik}=t_{ik}, \varepsilon_{ij}=1, \varepsilon_{ik}=2| T_{i0}=t_{i0}, \varepsilon_{i0}=2)^{ W^{1}_{1,2,2}}  \ * \\
    P(T_{ij}>t_{ij}, T_{ik}=t_{ik}, \varepsilon_{ij}=2, \varepsilon_{ik}=2| T_{i0}=t_{i0}, \varepsilon_{i0}=2)^{W^{2}_{2,2,2}}
    \end{align*}
    with
    \begin{align*}
        &P(T_{ij}>t_{ij}, T_{ik}=t_{ik}, \varepsilon_{ij}=1, \varepsilon_{ik}=2| T_{i0}=t_{i0}, \varepsilon_{i0}=2)\nonumber \\
    &=\dfrac{P(T_{ij}>t_{ij}, T_{ik}=t_{ik}, \varepsilon_{ij}=1, \varepsilon_{ik}=2,T_{i0}=t_{i0}, \varepsilon_{i0}=2)}{P(T_{i0}=t_{i0}, \varepsilon_{i0}=2)})\nonumber \\
    &=\dfrac{P(T_{ij}>t_{ij}, T_{ik}=t_{ik}, T_{i0}=t_{i0}| \varepsilon_{ij}=1, \varepsilon_{ik}=2, \varepsilon_{i0}=2)P(\varepsilon_{ij}=1, \varepsilon_{ik}=2,\varepsilon_{i0}=2)}{P(T_{i0}=t_{i0}, \varepsilon_{i0}=2)})\nonumber \\
    &=D_{\alpha_{21}}^{011}\{S_1(t_{ij}), S_2(t_{ik}), S_2(t_{i0})\}f_2(t_{ik})F_1(\infty|X_{ij})\nonumber\\
    &\propto D_{\alpha_{21}}^{011}\{S_1(t_{ij}), S_2(t_{ik}), S_2(t_{i0})\} \ \text{and},
    \end{align*}
    
    \begin{align*}
    &P(T_{ij}>t_{ij}, T_{ik}=t_{ik}, \varepsilon_{ij}=2, \varepsilon_{ik}=2| T_{i0}=t_{i0}, \varepsilon_{i0}=2)\nonumber \\
    &=\dfrac{P(T_{ij}>t_{ij}, T_{ik}=t_{ik}, \varepsilon_{ij}=2, \varepsilon_{ik}=2,T_{i0}=t_{i0}, \varepsilon_{i0}=2)}{P(T_{i0}=t_{i0}, \varepsilon_{i0}=2)})\nonumber \\
    &=\dfrac{P(T_{ij}>t_{ij}, T_{ik}=t_{ik}, T_{i0}=t_{i0}| \varepsilon_{ij}=2, \varepsilon_{ik}=2, \varepsilon_{i0}=2)P(\varepsilon_{ij}=2, \varepsilon_{ik}=2,\varepsilon_{i0}=2)}{P(T_{i0}=t_{i0}, \varepsilon_{i0}=2)})\nonumber \\
    &=D_{\alpha_{21}}^{011}\{S_2(t_{ij}), S_2(t_{ik}), S_2(t_{i0})\}f_2(t_{ik})F_2(\infty|X_{ij})\nonumber\\
    &\propto D_{\alpha_{21}}^{011}\{S_2(t_{ij}), S_2(t_{ik}), S_2(t_{i0})\}
    \end{align*}
\par
    We express:
    \begin{align*}
       W^{l}_{l,l',l_0}&=P(\varepsilon_{ij}=l| T_{ij}>t_{ij}, T_{ik}=t_{ik}, \varepsilon_{ik}=l', T_{i0}=t_{i0}, \varepsilon_{i0}=l_0)\\
       &=\dfrac{P(\varepsilon_{ij}=l, T_{ij}>t_{ij}, T_{ik}=t_{ik}, \varepsilon_{ik}=l', T_{i0}=t_{i0}, \varepsilon_{i0}=l_0)}{P(T_{ij}>t_{ij}, T_{ik}=t_{ik}, \varepsilon_{ik}=l', T_{i0}=t_{i0}, \varepsilon_{i0}=l_0)}\\
       &=\dfrac{P(T_{ij}>t_{ij}, T_{ik}=t_{ik}, T_{i0}=t_{i0}|\varepsilon_{ij}=l, \varepsilon_{ik}=l', \varepsilon_{i0}=l_0)P(\varepsilon_{ij}=l, \varepsilon_{ik}=l', \varepsilon_{i0}=l_0)}{\sum_{r=1}^{2}P(T_{ij}>t_{ij}, T_{ik}=t_{ik},  T_{i0}=t_{i0}, \varepsilon_{ik}=l', \varepsilon_{i0}=l_0, \varepsilon_{ij}=r)}\\
     &=\dfrac{P(T_{ij}>t_{ij}, T_{ik}=t_{ik}, T_{i0}=t_{i0}|\varepsilon_{ij}=l, \varepsilon_{ik}=l', \varepsilon_{i0}=l_0)P(\varepsilon_{ij}=l, \varepsilon_{ik}=l', \varepsilon_{i0}=l_0)}{\sum_{r=1}^{2}P(T_{ij}>t_{ij}, T_{ik}=t_{ik}, T_{i0}=t_{i0}|\varepsilon_{ij}=r, \varepsilon_{ik}=l', \varepsilon_{i0}=l_0)P(\varepsilon_{ij}=r, \varepsilon_{ik}=l', \varepsilon_{i0}=l_0)}
   \end{align*}
%%%%%%%%%%%%%%%%%%%%%%%%%%%%%%%%%%%%%%%%%%%%%%%%%%%%%%%%%%%%%%%%%%%%%%%%%%%%%%%%%%%%%%%%%%%%%
When, for each pair, both subjects have not observed any events, we distinguish 2 contributions that are weighted by \par $W^{ll'}_{l,l',l_0}=P(\varepsilon_{ij}=l,\varepsilon_{ik}=l'| T_{ij}>t_{ij}, T_{ik}=t_{ik}, T_{i0}=t_{i0}, \varepsilon_{i0}=l_0)$\\
 %%% 17 %%
 \item $\delta_{ij}=0$, $\delta_{ik}=0$, $\delta_{i0}=1$
 \begin{align*}
   P(T_{ij}>t_{ij}, T_{ik}>t_{ik}, \varepsilon_{ij}=1, \varepsilon_{ik}=1| T_{i0}=t_{i0}, \varepsilon_{i0}=1)^{W^{11}_{1,1,1}}  *\\
   P(T_{ij}>t_{ij}, T_{ik}>t_{ik}, \varepsilon_{ij}=2, \varepsilon_{ik}=1| T_{i0}=t_{i0}, \varepsilon_{i0}=1)^{W^{21}_{2,1,1}} *\\
   P(T_{ij}>t_{ij}, T_{ik}>t_{ik}, \varepsilon_{ij}=1, \varepsilon_{ik}=2| T_{i0}=t_{i0}, \varepsilon_{i0}=1)^{W^{12}_{1,2,1}} *\\
   P(T_{ij}>t_{ij}, T_{ik}>t_{ik}, \varepsilon_{ij}=2, \varepsilon_{ik}=2| T_{i0}=t_{i0}, \varepsilon_{i0}=1)^{W^{22}_{2,2,1}}
   \end{align*}
 with
    \begin{align*}
        &P(T_{ij}>t_{ij}, T_{ik}>t_{ik}, \varepsilon_{ij}=1, \varepsilon_{ik}=1| T_{i0}=t_{i0}, \varepsilon_{i0}=1)\nonumber \\
    &=\dfrac{P(T_{ij}>t_{ij}, T_{ik}>t_{ik}, \varepsilon_{ij}=1, \varepsilon_{ik}=1,T_{i0}=t_{i0}, \varepsilon_{i0}=1)}{P(T_{i0}=t_{i0}, \varepsilon_{i0}=1)})\nonumber \\
    &=\dfrac{P(T_{ij}>t_{ij}, T_{ik}>t_{ik}, T_{i0}=t_{i0}| \varepsilon_{ij}=1, \varepsilon_{ik}=1, \varepsilon_{i0}=1)P(\varepsilon_{ij}=1, \varepsilon_{ik}=1,\varepsilon_{i0}=1)}{P(T_{i0}=t_{i0}, \varepsilon_{i0}=1)})\nonumber \\
    &=D_{\alpha_{11}}^{001}\{S_1(t_{ij}), S_1(t_{ik}), S_1(t_{i0})\}F_1(\infty|X_{ij})F_1(\infty|X_{ik})\nonumber \\
    &\propto D_{\alpha_{11}}^{001}\{S_1(t_{ij}), S_1(t_{ik}), S_1(t_{i0})\}
   \\
 &\text{and,}\nonumber\\
    &P(T_{ij}>t_{ij}, T_{ik}>t_{ik}, \varepsilon_{ij}=2, \varepsilon_{ik}=1| T_{i0}=t_{i0}, \varepsilon_{i0}=1)\nonumber \\
    &=\dfrac{P(T_{ij}>t_{ij}, T_{ik}>t_{ik}, \varepsilon_{ij}=2, \varepsilon_{ik}=1,T_{i0}=t_{i0}, \varepsilon_{i0}=1)}{P(T_{i0}=t_{i0}, \varepsilon_{i0}=1)})\nonumber \\
    &=\dfrac{P(T_{ij}>t_{ij}, T_{ik}>t_{ik}, T_{i0}=t_{i0}| \varepsilon_{ij}=2, \varepsilon_{ik}=1, \varepsilon_{i0}=1)P(\varepsilon_{ij}=2, \varepsilon_{ik}=1,\varepsilon_{i0}=1)}{P(T_{i0}=t_{i0}, \varepsilon_{i0}=1)})\nonumber \\
    &=D_{\alpha_{21}}^{001}\{S_2(t_{ij}), S_1(t_{ik}), S_1(t_{i0})\}F_2(\infty|X_{ij})F_1(\infty|X_{ik})\nonumber \\
    &\propto D_{\alpha_{21}}^{001}\{S_2(t_{ij}), S_1(t_{ik}), S_1(t_{i0})\}
    \\
 &\text{and,}\nonumber\\
    &P(T_{ij}>t_{ij}, T_{ik}>t_{ik}, \varepsilon_{ij}=1, \varepsilon_{ik}=2| T_{i0}=t_{i0}, \varepsilon_{i0}=1)\nonumber \\
    &=\dfrac{P(T_{ij}>t_{ij}, T_{ik}>t_{ik}, \varepsilon_{ij}=1, \varepsilon_{ik}=2,T_{i0}=t_{i0}, \varepsilon_{i0}=1)}{P(T_{i0}=t_{i0}, \varepsilon_{i0}=1)})\nonumber \\
    &=\dfrac{P(T_{ij}>t_{ij}, T_{ik}>t_{ik}, T_{i0}=t_{i0}| \varepsilon_{ij}=1, \varepsilon_{ik}=2, \varepsilon_{i0}=1)P(\varepsilon_{ij}=1, \varepsilon_{ik}=2,\varepsilon_{i0}=1)}{P(T_{i0}=t_{i0}, \varepsilon_{i0}=1)})\nonumber \\
    &=D_{\alpha_{21}}^{001}\{S_1(t_{ij}), S_2(t_{ik}), S_1(t_{i0})\}F_2(\infty|X_{ij})F_2(\infty|X_{ik})\nonumber \\
    &\propto D_{\alpha_{21}}^{001}\{S_1(t_{ij}), S_2(t_{ik}), S_1(t_{i0})\}
   \\
 &\text{and,}\nonumber\\
    &P(T_{ij}>t_{ij}, T_{ik}>t_{ik}, \varepsilon_{ij}=2, \varepsilon_{ik}=2| T_{i0}=t_{i0}, \varepsilon_{i0}=1)\nonumber \\
    &=\dfrac{P(T_{ij}>t_{ij}, T_{ik}>t_{ik}, T_{i0}=t_{i0}| \varepsilon_{ij}=2, \varepsilon_{ik}=2, \varepsilon_{i0}=1)P(\varepsilon_{ij}=2, \varepsilon_{ik}=2,\varepsilon_{i0}=1)}{P(T_{i0}=t_{i0}, \varepsilon_{i0}=1)})\nonumber \\
    &=D_{\alpha_{21}}^{001}\{S_2(t_{ij}), S_2(t_{ik}), S_1(t_{i0})\}F_2(\infty|X_{ij})F_2(\infty|X_{ik})\nonumber \\
    &\propto D_{\alpha_{21}}^{001}\{S_2(t_{ij}), S_2(t_{ik}), S_1(t_{i0})\}
    \end{align*}
    
    %%%%%%%%%%%%%%%%%%%%%%%%%%%%%%%%%%%%%%%%%%%%%%%%%%%%%%%%%%%%%%%%%%%%%%%%%%%%%%%%%%%%%%%%%
    
 %%% 18 %%
\item $\delta_{ij}=0$, $\delta_{ik}=0$, $\delta_{i0}=2$
\begin{align*}
P(T_{ij}>t_{ij}, T_{ik}>t_{ik}, \varepsilon_{ij}=1, \varepsilon_{ik}=1| T_{i0}=t_{i0}, \varepsilon_{i0}=2)^{ W^{11}_{1,1,2}}  *\\
P(T_{ij}>t_{ij}, T_{ik}>t_{ik}, \varepsilon_{ij}=2, \varepsilon_{ik}=1| T_{i0}=t_{i0}, \varepsilon_{i0}=2)^{W^{21}_{2,1,2}} *\\
P(T_{ij}>t_{ij}, T_{ik}>t_{ik}, \varepsilon_{ij}=1, \varepsilon_{ik}=2| T_{i0}=t_{i0}, \varepsilon_{i0}=2)^{W^{12}_{1,2,2}} *\\
P(T_{ij}>t_{ij}, T_{ik}>t_{ik}, \varepsilon_{ij}=2, \varepsilon_{ik}=2| T_{i0}=t_{i0}, \varepsilon_{i0}=2)^{W^{22}_{2,2,2}}
\end{align*}
with
\begin{align*}
&P(T_{ij}>t_{ij}, T_{ik}>t_{ik}, \varepsilon_{ij}=1, \varepsilon_{ik}=1| T_{i0}=t_{i0}, \varepsilon_{i0}=2)\nonumber \\
&=\dfrac{P(T_{ij}>t_{ij}, T_{ik}>t_{ik}, \varepsilon_{ij}=1, \varepsilon_{ik}=1,T_{i0}=t_{i0}, \varepsilon_{i0}=2)}{P(T_{i0}=t_{i0}, \varepsilon_{i0}=2)})\nonumber \\
&=\dfrac{P(T_{ij}>t_{ij}, T_{ik}>t_{ik}, T_{i0}=t_{i0}| \varepsilon_{ij}=1, \varepsilon_{ik}=1, \varepsilon_{i0}=2)P(\varepsilon_{ij}=1, \varepsilon_{ik}=1,\varepsilon_{i0}=2)}{P(T_{i0}=t_{i0}, \varepsilon_{i0}=2)})\nonumber \\
&=D_{\alpha_{12}}^{001}\{S_1(t_{ij}), S_1(t_{ik}), S_2(t_{i0})\}F_1(\infty|X_{ij})F_1(\infty|X_{ik})\nonumber \\
&\propto D_{\alpha_{12}}^{001}\{S_1(t_{ij}), S_1(t_{ik}), S_2(t_{i0})\}\ \text{and,}\nonumber\\
&P(T_{ij}>t_{ij}, T_{ik}>t_{ik}, \varepsilon_{ij}=2, \varepsilon_{ik}=1| T_{i0}=t_{i0}, \varepsilon_{i0}=2)\nonumber \\
&=\dfrac{P(T_{ij}>t_{ij}, T_{ik}>t_{ik}, \varepsilon_{ij}=2, \varepsilon_{ik}=1,T_{i0}=t_{i0}, \varepsilon_{i0}=2)}{P(T_{i0}=t_{i0}, \varepsilon_{i0}=1)})\nonumber \\
&=\dfrac{P(T_{ij}>t_{ij}, T_{ik}>t_{ik}, T_{i0}=t_{i0}| \varepsilon_{ij}=2, \varepsilon_{ik}=1, \varepsilon_{i0}=1)P(\varepsilon_{ij}=2, \varepsilon_{ik}=1,\varepsilon_{i0}=2)}{P(T_{i0}=t_{i0}, \varepsilon_{i0}=2)})\nonumber \\
&=D_{\alpha_{21}}^{001}\{S_2(t_{ij}), S_1(t_{ik}), S_2(t_{i0})\}F_2(\infty|X_{ij})F_1(\infty|X_{ik})\nonumber \\
&\propto D_{\alpha_{21}}^{001}\{S_2(t_{ij}), S_1(t_{ik}), S_2(t_{i0})\}
 \  \text{and,}\nonumber\\
&P(T_{ij}>t_{ij}, T_{ik}>t_{ik}, \varepsilon_{ij}=1, \varepsilon_{ik}=2| T_{i0}=t_{i0}, \varepsilon_{i0}=2)\nonumber \\
&=\dfrac{P(T_{ij}>t_{ij}, T_{ik}>t_{ik}, \varepsilon_{ij}=1, \varepsilon_{ik}=2,T_{i0}=t_{i0}, \varepsilon_{i0}=2)}{P(T_{i0}=t_{i0}, \varepsilon_{i0}=2)})\nonumber \\
&=\dfrac{P(T_{ij}>t_{ij}, T_{ik}>t_{ik}, T_{i0}=t_{i0}| \varepsilon_{ij}=1, \varepsilon_{ik}=2, \varepsilon_{i0}=2)P(\varepsilon_{ij}=1, \varepsilon_{ik}=2,\varepsilon_{i0}=2)}{P(T_{i0}=t_{i0}, \varepsilon_{i0}=2)})\nonumber \\
&=D_{\alpha_{12}}^{001}\{S_1(t_{ij}), S_2(t_{ik}), S_2(t_{i0})\}F_1(\infty|X_{ij})F_2(\infty|X_{ik})\nonumber \\
&\propto D_{\alpha_{12}}^{001}\{S_1(t_{ij}), S_2(t_{ik}), S_2(t_{i0})\}
   \  \text{and,}\nonumber\\
&P(T_{ij}>t_{ij}, T_{ik}>t_{ik}, \varepsilon_{ij}=2, \varepsilon_{ik}=2| T_{i0}=t_{i0}, \varepsilon_{i0}=2)\nonumber \\
&=\dfrac{P(T_{ij}>t_{ij}, T_{ik}>t_{ik}, \varepsilon_{ij}=2, \varepsilon_{ik}=2,T_{i0}=t_{i0}, \varepsilon_{i0}=2)}{P(T_{i0}=t_{i0}, \varepsilon_{i0}=2)})\nonumber \\
&=\dfrac{P(T_{ij}>t_{ij}, T_{ik}>t_{ik}, T_{i0}=t_{i0}| \varepsilon_{ij}=2, \varepsilon_{ik}=2, \varepsilon_{i0}=2)P(\varepsilon_{ij}=2, \varepsilon_{ik}=2,\varepsilon_{i0}=2)}{P(T_{i0}=t_{i0}, \varepsilon_{i0}=2)})\nonumber \\
&=D_{\alpha_{22}}^{001}\{S_2(t_{ij}), S_2(t_{ik}), S_2(t_{i0})\}F_2(\infty|X_{ij})F_2(\infty|X_{ik})\nonumber \\
&\propto D_{\alpha_{22}}^{001}\{S_2(t_{ij}), S_2(t_{ik}), S_2(t_{i0})\}
    \end{align*}

    We express:
    \begin{align*}
       W^{ll'}_{l,l',l_0}&=P(\varepsilon_{ij}=l, \varepsilon_{ik}=l'| T_{ij}>t_{ij}, T_{ik}>t_{ik}, T_{i0}=t_{i0}, \varepsilon_{i0}=l_0)\\
       &=\dfrac{P(\varepsilon_{ij}=l, T_{ij}>t_{ij}, T_{ik}>t_{ik}, \varepsilon_{ik}=l', T_{i0}=t_{i0}, \varepsilon_{i0}=l_0)}{P(T_{ij}>t_{ij}, T_{ik}>t_{ik}, T_{i0}=t_{i0}, \varepsilon_{i0}=l_0)}\\
       &=\dfrac{P(T_{ij}>t_{ij}, T_{ik}>t_{ik}, T_{i0}=t_{i0}|\varepsilon_{ij}=l, \varepsilon_{ik}=l', \varepsilon_{i0}=l_0)P(\varepsilon_{ij}=l, \varepsilon_{ik}=l', \varepsilon_{i0}=l_0)}{\sum_{s=1}^{2}\sum_{r=1}^{2}P(T_{ij}>t_{ij}, T_{ik}>t_{ik},  T_{i0}=t_{i0}, \varepsilon_{ij}=r, \varepsilon_{ik}=s, \varepsilon_{i0}=l_0)}\\
     &=\resizebox{1.05\linewidth}{!}{$\dfrac{P(T_{ij}>t_{ij}, T_{ik}>t_{ik}, T_{i0}=t_{i0}|\varepsilon_{ij}=l, \varepsilon_{ik}=l', \varepsilon_{i0}=l_0)P(\varepsilon_{ij}=l, \varepsilon_{ik}=l', \varepsilon_{i0}=l_0)}{\sum_{s=1}^{2}\sum_{r=1}^{2}P(T_{ij}>t_{ij}, T_{ik}>t_{ik}, T_{i0}=t_{i0}|\varepsilon_{ij}=r, \varepsilon_{ik}=s, \varepsilon_{i0}=l_0)P(\varepsilon_{ij}=r, \varepsilon_{ik}=s, \varepsilon_{i0}=l_0)}$}
   \end{align*}
\end{enumerate}

\subsection{Appendix B}
%%%%%%%%%%%%%%%%%%%%%%%%%%%%%%%%%%%%%%%%%%%%%%%%%%%%%%%%%%%%
\begin{align*}
R(t)
&=
\frac{P(E_{j}=1 \mid T_{j}=t,X_{j})}{P(E_{j}=1 \mid T_{j}>t,X_{j})}
\times
\frac{P(E_{j}=2 \mid T_{j}>t,X_{j})}{P(E_{j}=2 \mid T_{j}=t,X_{j})}
\\[0.3cm]
&=
\frac{
\dfrac{f_1(t\mid X_{j})}{f_1(t\mid X_{j})+f_2(t\mid X_{j})}
}{
\dfrac{F_1(\infty\mid X_{j})-F_1(t\mid X_{j})}{S(t\mid X_{j})}
}
\times
\frac{
\dfrac{F_2(\infty\mid X_{j})-F_2(t\mid X_{j})}{S(t\mid X_{j})}
}{
\dfrac{f_2(t\mid X_{j})}{f_1(t\mid X_{j})+f_2(t\mid X_{j})}
}
\\[0.3cm]
&=
\frac{
f_1(t\mid X_{j})\left\{F_2(\infty\mid X_{j})-F_2(t\mid X_{j})\right\}
}{
f_2(t\mid X_{j})\left\{F_1(\infty\mid X_{j})-F_1(t\mid X_{j})\right\}
}.
\end{align*}
%%%%%%%%%%%%%%%%%%%%%%%%%%%%%%%%%%%%%%%%%%%%%%%%%%%%%%%%%%%%%%%%%%%%%5
\subsection{Appendix C}
\begin{align*}
\pi_{j,l,k,l'}(t,s)
=
\frac{
\mathrm{ODDS}\big(T_j \le t, E_j = l \mid T_k \le s, E_k = l', X_j, X_k\big)
}{
\mathrm{ODDS}\big(T_j \le t, E_j = l \mid X_j\big)
}.
\end{align*}
where, for any events $A$ and $B$, $\mathrm{ODDS}(A) = \frac{P(A)}{1 - P(A)}$ and $\mathrm{ODDS}(A \mid B) = \frac{P(A \mid B)}{1 - P(A \mid B)}.$ 

\noindent Rewriting this expression in terms of probabilities yields
\[
\pi_{j,l,k,l'}(t,s)=\frac{Q_{jl,kl'}(t,s|X_j, X_k)}{F_{l'}(s| X_k) - Q_{jl,kl'}(t,s|X_j, X_k)}\cdot\frac{1 - F_{l}(t|X_j)}{F_{l}(t|X_j)},
\]
where 
\begin{align*} 
Q_{jl,kl'}(t,s|X_j, X_k)&=P(T_j \le t, E_j = l, T_k \le s, E_k = l'|X_j, X_k)\\
&=\Big[1 - S_l(t|X_j) - S_{l'}(s|X_k)+ C_{ll'}\big(S_l(t|X_j), S_{l'}(s| X_k)\big)\Big]F_{l}(\infty|X_j)\,F_{l'}(\infty| X_k).
\end{align*}

%%%%%%%%%%%%%%%%%%%%%%%%%%%%%%%%%%%%%%%%%%%%%%%5
\subsection{Appendix D}
\begin{align*}
\mathrm{OR}_{jk}(t,s)
&=
\frac{
P(E_j = 1, E_k = 1 \mid T_j = t, T_k = s, X_j, X_k)\;
P(E_j = 2, E_k = 2 \mid T_j = t, T_k = s, X_j, X_k)
}{
P(E_j = 1, E_k = 2 \mid T_j = t, T_k = s, X_j, X_k)\;
P(E_j = 2, E_k = 1 \mid T_j = t, T_k = s, X_j, X_k)
}\\
&=
\frac{
D_{11}\big(S_1(t \mid X_j), S_1(s \mid X_k)\big)\;
D_{11}\big(S_2(t \mid X_j), S_2(s \mid X_k)\big)
}{
D_{11}\big(S_1(t \mid X_j), S_2(s \mid X_k)\big)\;
D_{11}\big(S_2(t \mid X_j), S_1(s \mid X_k)\big)
},
\end{align*}
where $D_{11}(u,v) = \dfrac{\partial^2 D(u,v)}{\partial u \partial v}$ denotes the second partial derivative of the copula $D$, taken here as a bivariate Gaussian copula.

\subsection{Appendix E}
\begin{align*}
CR (t)&= P(T_j<t, E_j=1 \mid T_j>s,\; T_k=t_0,\; E_k=l)\\
&\vspace{0cm}\\
&=\frac{\{D_{01}\{S_1(s|X_j),S_l(t_0|X_k)\}
-
D_{01}\{S_1(t|X_j),S_l(t_0|X_k)\}\}F_{1}(\infty|X_j)F_{l}(\infty|X_k)
}{D_{01}\{S_1(s|X_j),S_l(t_0|X_k)\}F_{1}(\infty|X_j)F_{l}(\infty|X_k)
+D_{01}\{S_1(s|X_j),S_l(t_0|X_k)\}F_{2}(\infty|X_j) F_{l}(\infty|X_k)
}\\
&\vspace{0cm}\\
&=\frac{\{D_{01}\{S_1(s|X_j),S_l(t_0|X_k)\}
-
D_{01}\{S_1(t|X_j),S_l(t_0|X_k)\}\}F_{1}(\infty|X_j)
}{D_{01}\{S_1(s|X_j),S_l(t_0|X_k)\}F_{1}(\infty|X_j)
+D_{01}\{S_1(s|X_j),S_l(t_0|X_k)\}F_{2}(\infty|X_j)
},
\end{align*}
where $D_{01}(u,v) = \dfrac{\partial D(u,v)}{ \partial v}$ denotes the  partial derivative of the copula $D$, taken here as a bivariate Gaussian copula.

\subsection{Appendix F}
\begin{figure}[H]
    \centering 
        \includegraphics[scale=0.8]{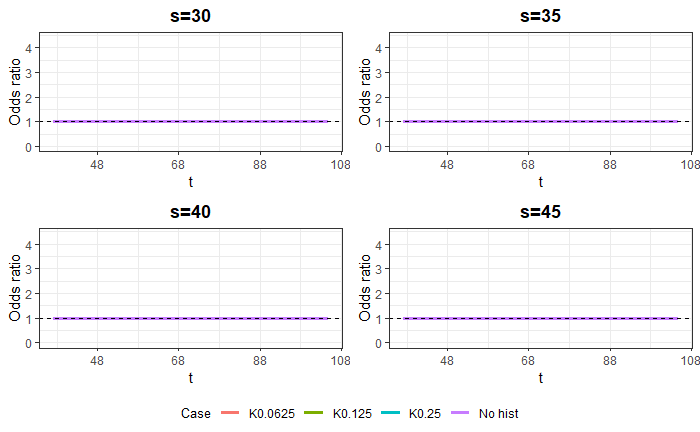}
         \caption{Estimated cumulative cross-odds ratio for breast cancer $(E_j=1, E_k=2)$ in BRCA1 carriers under different s}
         \label{hiera}
    \label{fig:my_label}
\end{figure}

\begin{figure}[H]
    \centering 
        \includegraphics[scale=0.8]{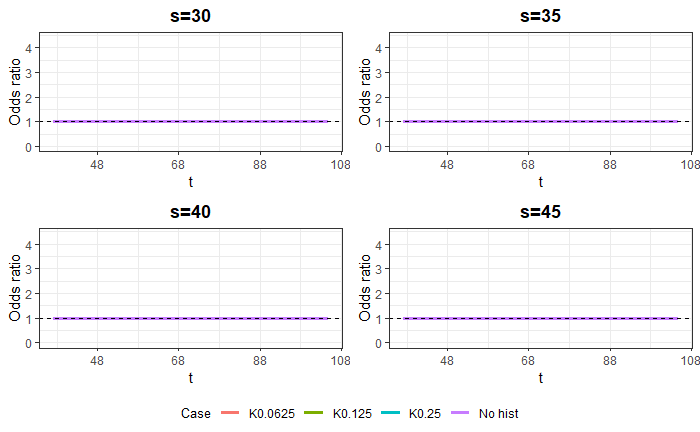}
         \caption{Estimated cumulative cross-odds ratio for ovarian cancer $(E_j=2, E_k=1)$ in BRCA1 carriers  under different s}
         \label{hiera}
    \label{fig:my_label}
\end{figure}

\begin{figure}[H]
    \centering 
        \includegraphics[scale=0.8]{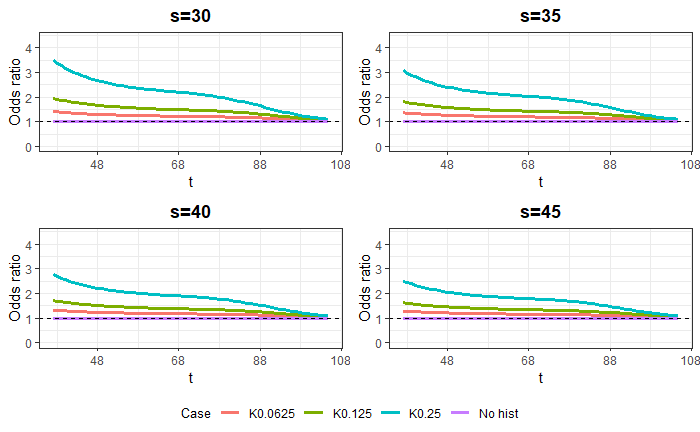}
         \caption{Estimated cumulative cross-odds ratio for breast cancer , $E_j=1, E_k=1$; BRCA1 carrier j and BRCA1 noncarrier k,  under different s.}
         \label{hiera}
    \label{fig:my_label}
\end{figure}

\begin{figure}[H]
    \centering 
        \includegraphics[scale=0.8]{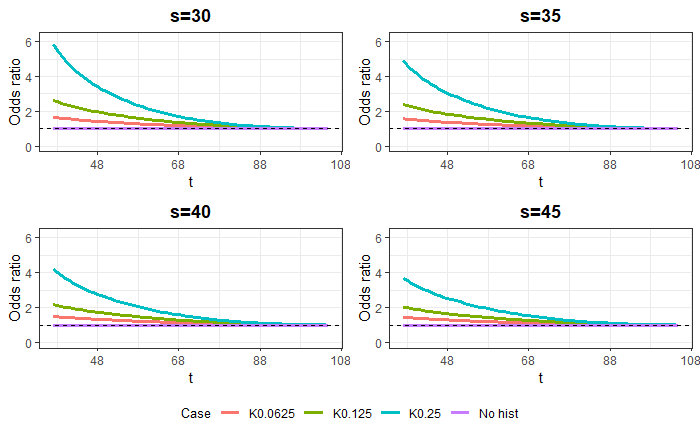}
         \caption{Estimated cumulative cross-odds ratio for ovarian cancer, $E_j=2, E_k=2$; BRCA1 carrier j and BRCA1 noncarrier k,  under different s.}
         \label{hiera}
    \label{fig:my_label}
\end{figure}

\begin{figure}[H]
    \centering 
        \includegraphics[scale=0.8]{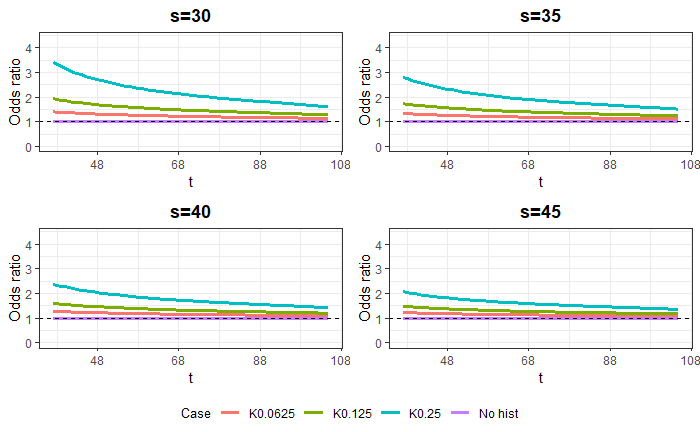}
         \caption{Estimated cumulative cross-odds ratio for breast cancer , $E_j=1, E_k=1$; BRCA1 noncarrier j and BRCA1 carrier k,  under different s.}
         \label{hiera}
    \label{fig:my_label}
\end{figure}

\begin{figure}[H]
    \centering 
        \includegraphics[scale=0.8]{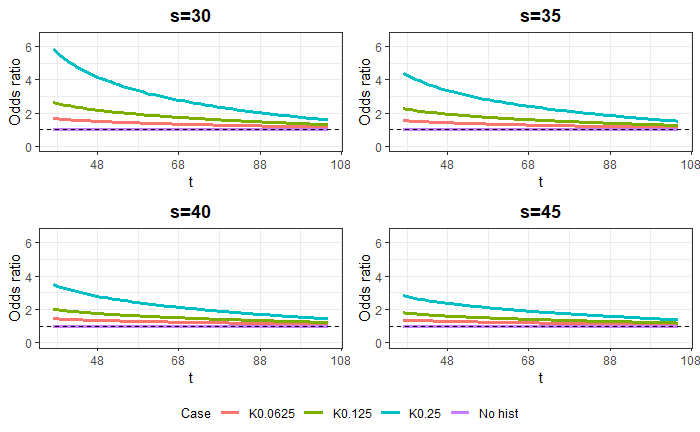}
         \caption{Estimated cumulative cross-odds ratio for ovarian cancer, $E_j=2, E_k=2$; BRCA1 noncarrier j and BRCA1 carrier k,  under different s.}
         \label{hiera}
    \label{fig:my_label}
\end{figure}

\begin{figure}[H]
    \centering 
        \includegraphics[scale=0.8]{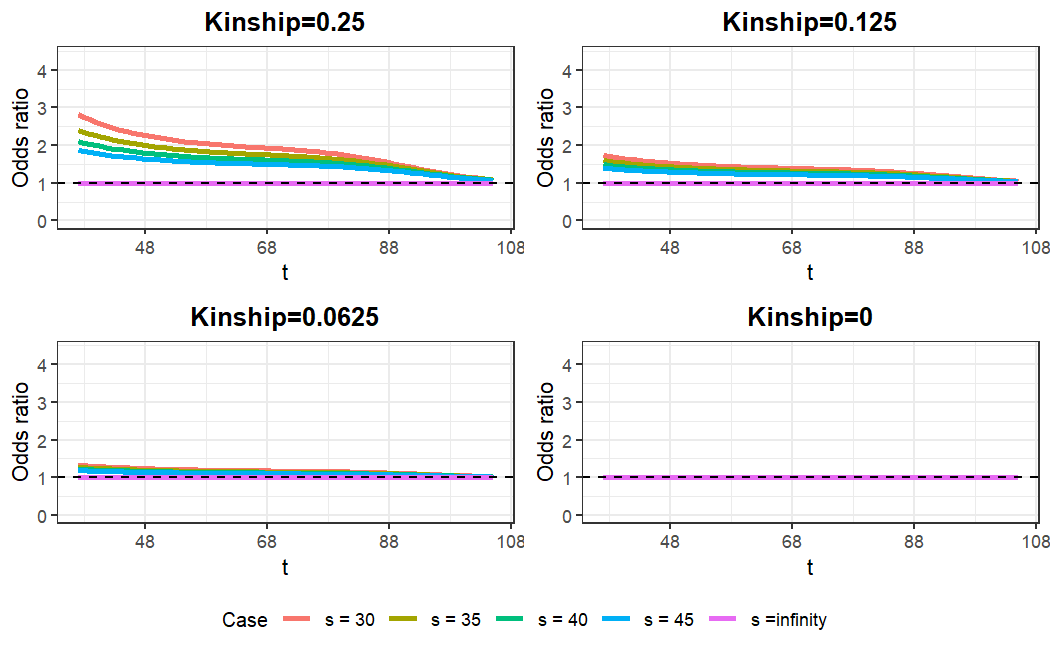}
         \caption{Estimated cumulative cross-odds ratio for breast cancer $(E_j=1, E_k=1)$ in BRCA1 carriers under different kinship}
         \label{hiera}
    \label{fig:my_label}
\end{figure}

\begin{figure}[H]
    \centering 
        \includegraphics[scale=0.8]{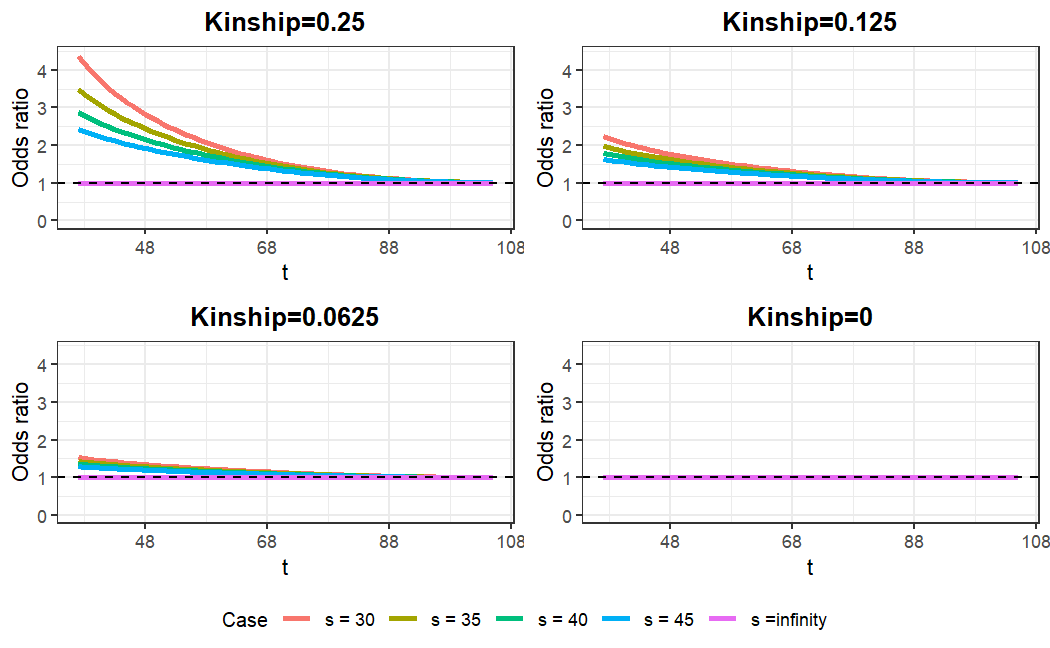}
         \caption{Estimated cumulative cross-odds ratio for ovarian cancer $(E_j=2, E_k=2)$ in BRCA1 carriers  under different kinship}
         \label{hiera}
    \label{fig:my_label}
\end{figure}

\noindent \textbf{ Somes 3D graphics }
%%%%%%%%%%%%%%%%%%%%%%%%%
\begin{figure}[H]
    \centering 
        \includegraphics[scale=0.4]{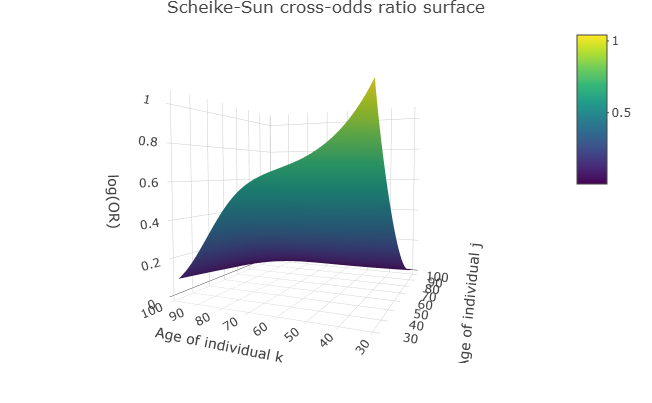}
         \caption{3D surface of the cumulative cross-odds ratio for breast cancer $(E_j=1, E_k=1)$ in BRCA1 carriers under kinship 0.25}
         \label{hiera}
    \label{fig:my_label}
\end{figure}

\begin{figure}[H]
    \centering 
        \includegraphics[scale=0.4]{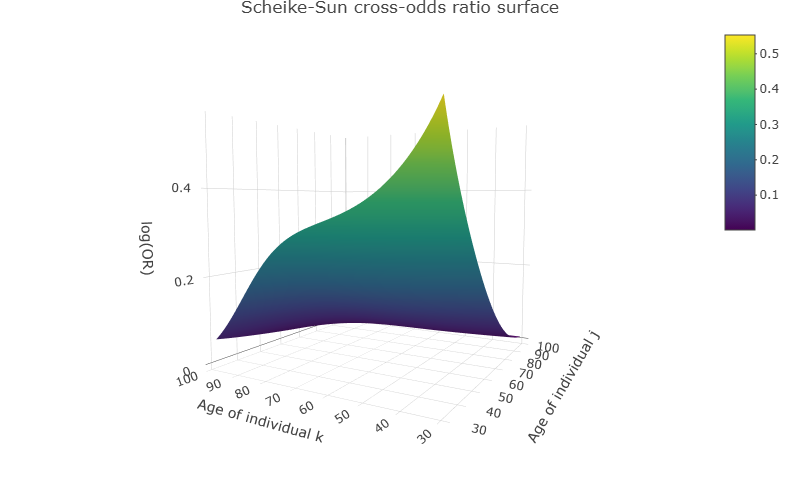}
         \caption{3D surface of the cumulative cross-odds ratio for breast cancer $(E_j=1, E_k=1)$ in BRCA1 carriers under kinship 0.125}
         \label{hiera}
    \label{fig:my_label}
\end{figure}

\begin{figure}[H]
    \centering 
        \includegraphics[scale=0.4]{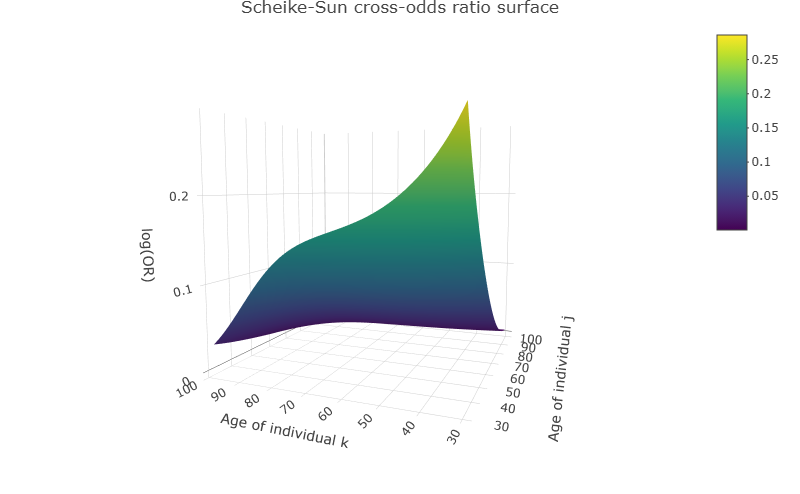}
         \caption{3D surface of the cumulative cross-odds ratio for breast cancer $(E_j=1, E_k=1)$ in BRCA1 carriers under kinship 0.0625}
         \label{hiera}
    \label{fig:my_label}
\end{figure}

%%%%%%%%%%%%%%%%%%%%%%ovaire%%%%%%%%%%%%%%%%%%%%%

\begin{figure}[H]
    \centering 
        \includegraphics[scale=0.4]{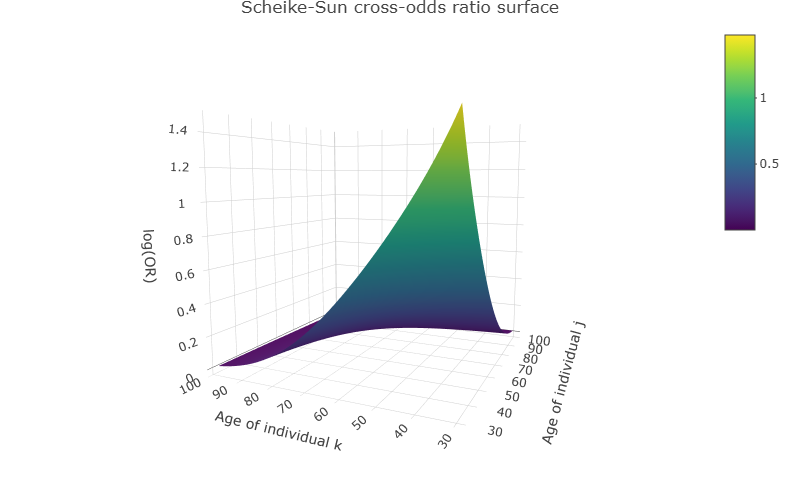}
         \caption{3D surface of the cumulative cross-odds ratio for ovarian cancer $(E_j=2, E_k=2)$ in BRCA1 carriers under kinship 0.25}
         \label{hiera}
    \label{fig:my_label}
\end{figure}

\begin{figure}[H]
    \centering 
        \includegraphics[scale=0.4]{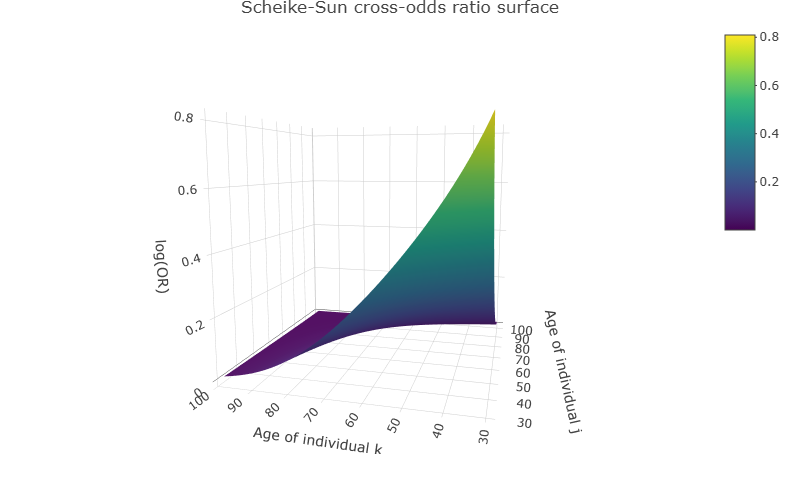}
         \caption{3D surface of the cumulative cross-odds ratio for ovarian cancer $(E_j=2, E_k=2)$ in BRCA1 carriers under kinship 0.125}
         \label{hiera}
    \label{fig:my_label}
\end{figure}

\begin{figure}[H]
    \centering 
        \includegraphics[scale=0.4]{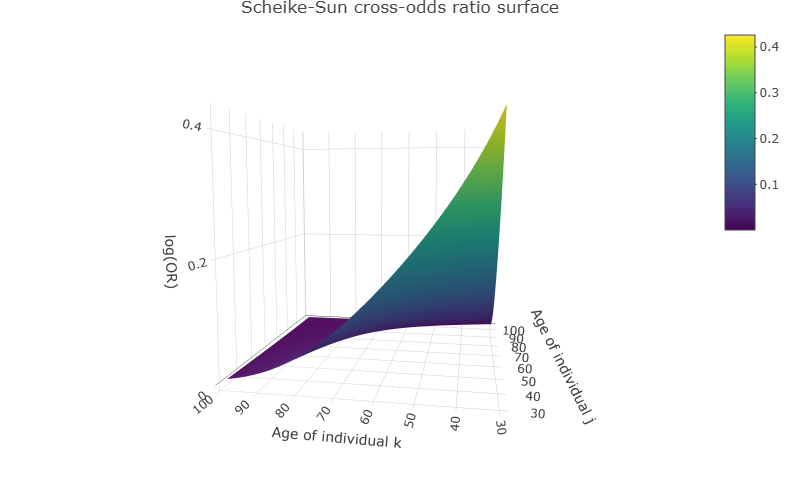}
         \caption{3D surface of the cumulative cross-odds ratio for ovarian cancer  $(E_j=2, E_k=2)$ in BRCA1 carriers under kinship 0.0625}
         \label{hiera}
    \label{fig:my_label}
\end{figure}

\subsection{Appendix G}

\begin{figure}[H]
    \centering 
        \includegraphics[scale=0.8]{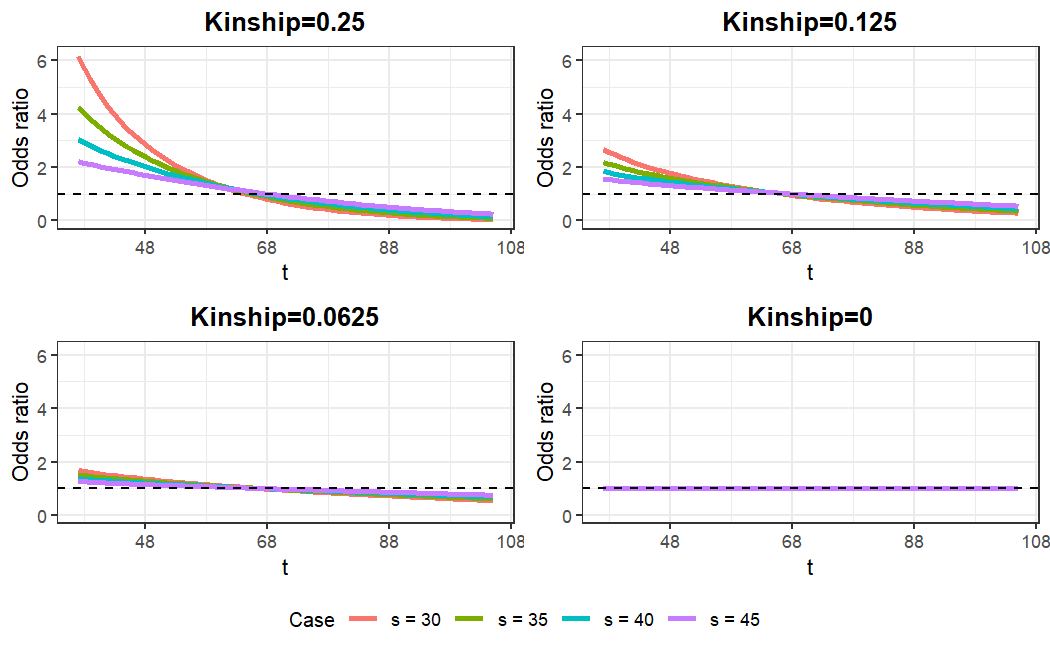}
 % \caption{Estimated instantaneous cross-odds ratio for identical cancer types among BRCA1 carriers}
  \caption{Instantaneous cross-odds ratio $\mathrm{OR}_{jk}(t,s)$  among \emph{BRCA1} mutation carriers, for different values of the kinship coefficient.}
         \label{hiera}
    \label{fig:my_label}
\end{figure}

\noindent \textbf{ Somes 3D graphics }
\begin{figure}[H]
    \centering 
        \includegraphics[scale=0.5]{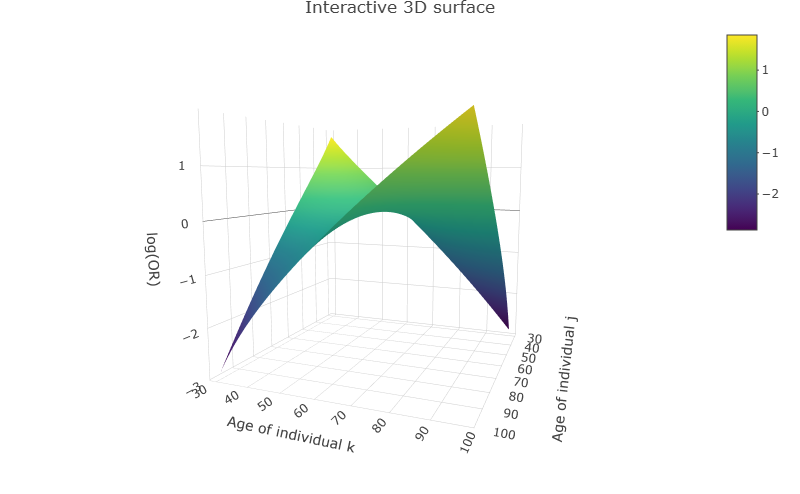}
         \caption{3D surface of the instantaneous cross-odds ratio under kinship 0.25}
         \label{hiera}
    \label{fig:my_label}
\end{figure}

\begin{figure}[H]
    \centering 
        \includegraphics[scale=0.5]{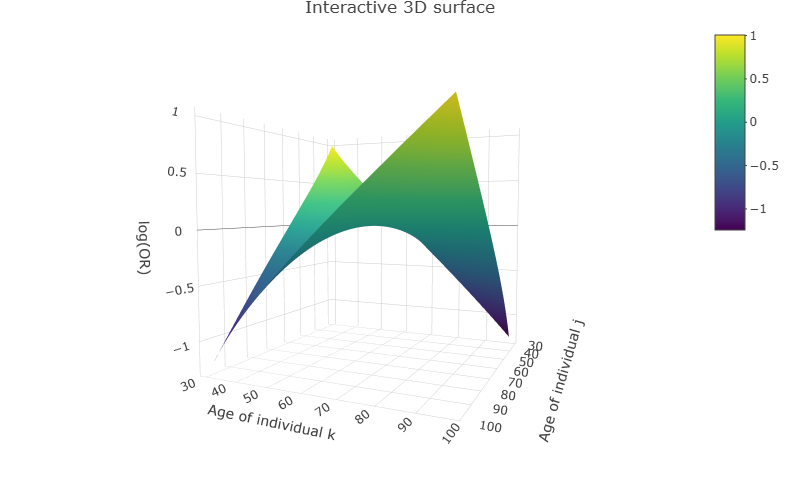}
         \caption{3D surface of the instantaneous cross-odds ratio under kinship 0.125}
         \label{hiera}
    \label{fig:my_label}
\end{figure}

\begin{figure}[H]
    \centering 
        \includegraphics[scale=0.5]{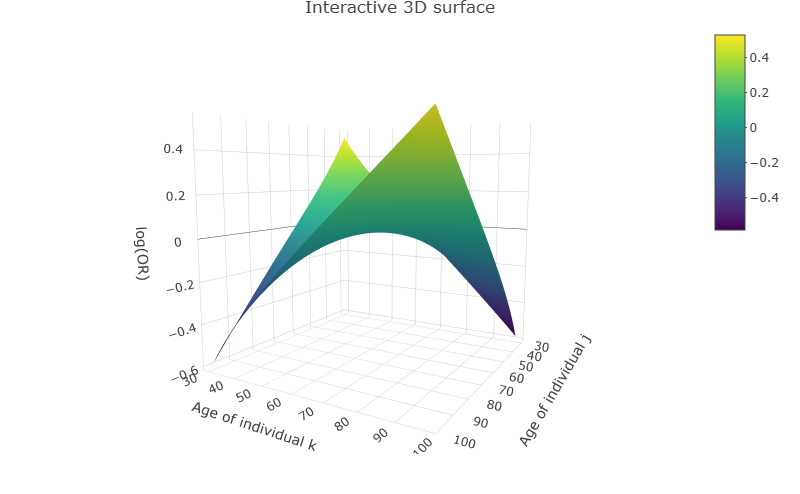}
         \caption{3D surface of the instantaneous cross-odds ratio under kinship 0.0625}
         \label{hiera}
    \label{fig:my_label}
\end{figure}

\begin{figure}[H]
    \centering 
        \includegraphics[scale=0.5]{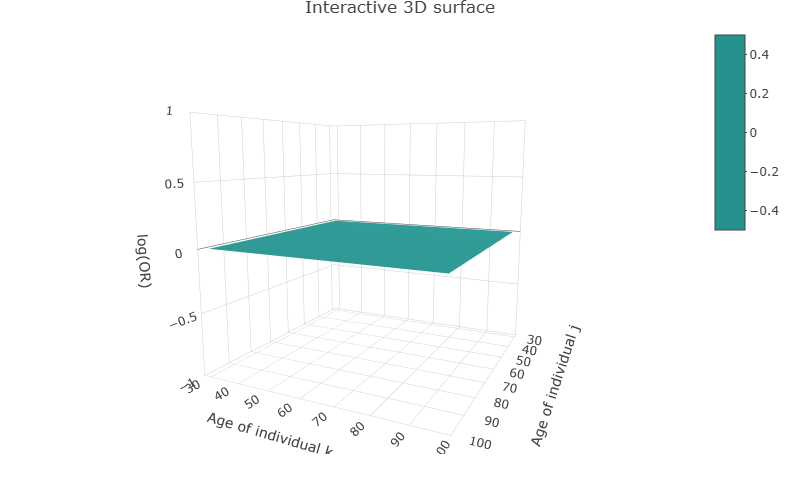}
         \caption{3D surface of the instantaneous cross-odds ratio under kinship 0}
         \label{hiera}
    \label{fig:my_label}
\end{figure}

\subsection{Appendix H}

\begin{figure}[H]
    \centering 
        \includegraphics[scale=0.9]{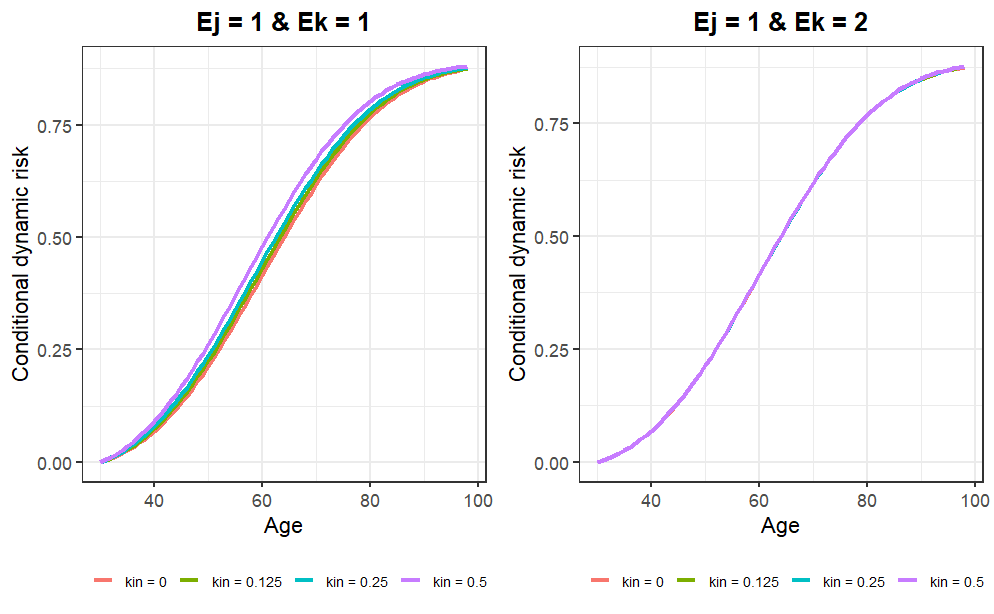}
               \caption{estimated conditional dynamic risk of breast cancer for different kinship coefficients ($0.25$, $0.125$, $0.0625$, and $0$) for BRCA1 carrier}
         \label{hiera}
    \label{fig:my_label}
\end{figure}

\subsection*{Appendix I}

\begin{figure}[H]
    \centering 
        \includegraphics[scale=0.8]{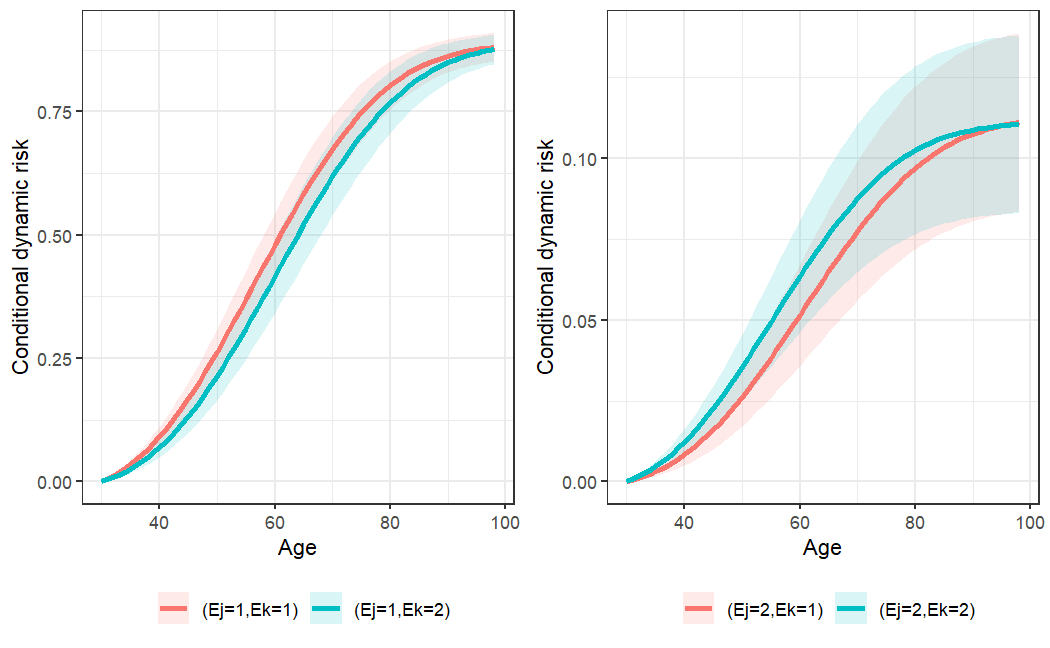}
         \caption{conditional dynamic risk of developing breast cancer (left panel) or ovarian cancer (righ panel) as a function of age, according to the kinship ($0.25$, $0.125$, $0.0625$, and $0$) coefficient with an affected relative and the type of cancer diagnosed in that relative. Red lines correspond to a relative diagnosed with breast cancer ($E_k=1$), blue lines to a relative diagnosed with ovarian cancer ($E_k=2$). All results are presented for $s=30$ and $t_0=45$, and BRCA1 carrier.}
         \label{hiera}
    \label{fig:my_label}
\end{figure}
\newpage
\subsection{Appendix J:  Database structure}
\begin{table}[htbp]
\centering
%\caption{Distribution of families according to family size}
\label{tab:family_size}
\begin{tabular}{ccc}
\toprule
\textbf{Family size} & \textbf{Number of families} & \textbf{Total individuals} \\
\midrule
1  & 37 & 37  \\
2  & 47 & 94  \\
3  & 57 & 171 \\
4  & 45 & 180 \\
5  & 56 & 280 \\
6  & 39 & 234 \\
7  & 31 & 217 \\
8  & 19 & 152 \\
9  & 9  & 81  \\
10 & 9  & 90  \\
11 & 10 & 110 \\
12 & 2  & 24  \\
13 & 5  & 65  \\
14 & 3  & 42  \\
15 & 2  & 30  \\
43 & 1  & 43  \\
77 & 1  & 77  \\
\midrule
\textbf{Total} & \textbf{373} & \textbf{1927} \\
\bottomrule
\end{tabular}
\caption{Distribution of families according to family size}
\end{table}

\subsection{Appendix K:  Simulation}

\noindent We conducted a simulation study to evaluate the finite-sample performance of the proposed estimation procedure in a familial competing-risks setting. We considered a setting with $I$ independent families having identical structures. Each family consisted of 6 individuals. Specifically, the family structure was composed of 3 pairs of siblings, and any two individuals who were not siblings were assumed to be cousins. Therefore, the kinship matrix was taken as
\[
\phi=
\begin{pmatrix}
0.5 & 0.25 & 0.0625 & 0.0625 & 0.0625 & 0.0625 \\
0.25 & 0.5 & 0.0625 & 0.0625 & 0.0625 & 0.0625 \\
0.0625 & 0.0625 & 0.5 & 0.25 & 0.0625 & 0.0625 \\
0.0625 & 0.0625 & 0.25 & 0.5 & 0.0625 & 0.0625 \\
0.0625 & 0.0625 & 0.0625 & 0.0625 & 0.5 & 0.25 \\
0.0625 & 0.0625 & 0.0625 & 0.0625 & 0.25 & 0.5
\end{pmatrix}.
\]
%%%%%%%%%%%%%%%%%%%%%%%

\noindent We considered two covariates, $X_{ij1}$ a binary variable such as $P(X_{ij1}=0)=P(X_{ij1}=1)=1/2$ and $X_{ij2}$ a uniform variable over $[0,4]$. Given these covariates, the hazard function of $T_{ij}$ is
$$\lambda(t|X_{ij1},X_{ij2})= \lambda_1(t|X_{ij1},X_{ij2}) + \lambda_2(t|X_{ij1},X_{ij2})\   \text{with}\  \lambda_l(t|X_{ij1},X_{ij2})=\lambda_{0l}(t) e^{\beta_{1l}X_{ij1} + \beta_{2l}X_{ij1} )}.$$ 

\noindent Here, we set $\lambda_{0l}(t)=\alpha_{l}\lambda_{l} t^{\alpha_l-1}$, $\lambda_1=\lambda_2=1/1000$, $\alpha_1=\alpha_2=4$, $\beta_{11}=\beta_{12}=1$ and $\beta_{21}=\beta_{22}=-1$.
\noindent The proband ages at entry $\{a_i, i=1,\cdots,I\}$ were generated from a Weibull distribution such that $E[a_i]=4$ and $Var[a_i]=2$, whereas the censoring times $\{c_{ij},i=1,\cdots,I, j=0\cdots,5\}$ were generated from a uniform distribution over $[5,10]$. Furthermore, we set the heritability parameters to $h_{11}=h_{22}=0.4$ and $h_{12}=h_{21}=0.2$.
%%%%%%%%%%%%%%%%%%%%%%%%%%%%%%%%%

\noindent For subject $j$ in family $i$, let $T_{ij}$ denote the latent time to the first event and let $\varepsilon_{ij}\in\{1,2\}$ denote the cause of the first event, where cause 1 and cause 2 represent the two competing cancer types. The observed data were
\[
Y_{ij}=\min(T_{ij},C_{ij}),
\qquad
\delta_{ij}=\mathbf{1}(T_{ij}<C_{ij})\times\varepsilon_{ij}\in\{0,1,2\} ,
\]
where $C_{ij}$ denotes the censoring time. Thus, $\varepsilon_{ij}$ was observed only when $\delta_{ij}\geq1$.

\noindent We considered two covariates, $X_{ij1}$ and $X_{ij2}$. The first covariate was binary, generated as
\[
P(X_{ij1}=1)=p,
\qquad
P(X_{ij1}=-1)=1-p,
\]
and the second covariate was generated from a uniform distribution over $[-1,1]$, i.e.
\[
X_{ij2}\sim \mathcal U(-1,1).
\]

\noindent For $l\in\{1,2\}$, let
\[
F_l(t\mid X_{ij1},X_{ij2})
=
P(T_{ij}\le t,\varepsilon_{ij}=l\mid X_{ij1},X_{ij2})
\]
denote the cumulative incidence function associated with cause $l$. In our simulation scheme, the cause of the first event was generated first, and the event time was then generated conditionally on that cause. For non-probands, the probability of cause 1 was defined as
\[
P(\varepsilon_{ij}=1\mid X_{ij1},X_{ij2})
=
F_1(\infty\mid X_{ij1},X_{ij2}),
\qquad j=1,\dots,5,
\]
whereas for the proband, conditionally on the age at entry $a_i$, we set
\[
P(\varepsilon_{i0}=1\mid T_{i0}\le a_i, X_{i01},X_{i02})
=
\frac{F_1(a_i\mid X_{i01},X_{i02})}
{F_1(a_i\mid X_{i01},X_{i02})+F_2(a_i\mid X_{i01},X_{i02})}.
\]

\noindent Conditional on $\varepsilon_{ij}=l$, the distribution function of the latent first-event time was
\[
P(T_{ij}\le t\mid \varepsilon_{ij}=l, X_{ij1},X_{ij2})
=
\frac{F_l(t\mid X_{ij1},X_{ij2})}
{F_l(\infty\mid X_{ij1},X_{ij2})},
\]
so that the corresponding survival function was
\[
P(T_{ij}>t\mid \varepsilon_{ij}=l, X_{ij1},X_{ij2})
=
1-\frac{F_l(t\mid X_{ij1},X_{ij2})}
{F_l(\infty\mid X_{ij1},X_{ij2})}.
\]

\noindent Familial dependence was introduced through a latent Gaussian representation. Let
\[
Z_i=(Z_{i0},\dots,Z_{i5})^\top \sim \mathcal N_6(0,V_i),
\]
where the covariance matrix $V_i$ depended on the kinship coefficients and the cause-specific dependence parameters $h_{11}$, $h_{22}$ and $h_{12}$. More precisely, for $j\neq k$,
\[
V_{i,jk}=
\begin{cases}
h_{11}\phi_{jk}, & \text{if } \varepsilon_{ij}=1 \text{ and } \varepsilon_{ik}=1,\\[4pt]
h_{22}\phi_{jk}, & \text{if } \varepsilon_{ij}=2 \text{ and } \varepsilon_{ik}=2,\\[4pt]
h_{12}\phi_{jk}, & \text{if } \varepsilon_{ij}\neq \varepsilon_{ik},
\end{cases}
\]
and $V_{i,jj}=1$ for all $j=0,\dots,5$.

\noindent Given the covariates, the age at entry, and the censoring times, we generated the competing-risks observations with ascertainment as follows.

\noindent For $i=1,\dots,I$:
\begin{enumerate}
\item[$\bullet$] Generate the proband covariates $(X_{i01},X_{i02})$ and the non-proband covariates $(X_{ij1},X_{ij2})$, $j=1,\dots,5$.
\item[$\bullet$] Generate the proband cause $\varepsilon_{i1}$ from a Bernoulli distribution with probability
\[
P(\varepsilon_{i0}=1\mid T_{i0}\le a_i, X_{i01},X_{i02})
=
\frac{F_1(a_i\mid X_{i01},X_{i02})}
{F_1(a_i\mid X_{i01},X_{i02})+F_2(a_i\mid X_{i01},X_{i02})}.
\]
\item[$\bullet$] For $j=1,\dots,5$, generate $\varepsilon_{ij}$ from a Bernoulli distribution with probability
\[
P(\varepsilon_{ij}=1\mid X_{ij1},X_{ij2})=F_1(\infty\mid X_{ij1},X_{ij2}).
\]
\item[$\bullet$] Conditional on $\varepsilon_{i0}=l$, generate $v_{i0}$ from a uniform distribution over $[0,1]$ and solve
\[
P(T_{i1}>t\mid T_{i1}\le a_i, \varepsilon_{i0}=l,X_{i01},X_{i02})=v_{i0}
\]
in $t$ to obtain
\[
1-\frac{F_l(t_{i0}\mid X_{i01},X_{i02})}
{F_l(a_i\mid X_{i01},X_{i02})}
=
v_{i0},
\qquad 0\le t_{i0}\le a_i.
\]
Equivalently,
\[
F_l(t_{i0}\mid X_{i01},X_{i02})
=
(1-v_{i0})F_k(a_i\mid X_{i01},X_{i02}).
\]
Because the cumulative incidence functions did not admit a closed-form inverse, $t_{i0}$ was obtained by numerical inversion.

\item[$\bullet$] Compute
\[
u_{i0}
=
P(T_{i0}>t_{i0}\mid \varepsilon_{i0},X_{i01},X_{i02})
=
1-\frac{F_{\varepsilon_{i0}}(t_{i0}\mid X_{i01},X_{i02})}
{F_{\varepsilon_{i0}}(\infty\mid X_{i01},X_{i02})},
\]
and define
\[
z_{i0}=\Phi^{-1}(u_{i0}).
\]

\item[$\bullet$] Construct the covariance matrix $V_i$ using the vector of causes
\[
(\varepsilon_{i0},\dots,\varepsilon_{i5})
\]
and generate $(Z_{i1},\dots,Z_{i5})^\top$ from the conditional multivariate normal distribution given $Z_{i0}=z_{i0}$.

\item[$\bullet$] For $j=1,\dots,5$:
\begin{enumerate}
\item[{\it (i)}] Compute
\[
u_{ij}=\Phi(Z_{ij}).
\]

\item[{\it (ii)}] Conditional on $\varepsilon_{ij}=l$, solve
\[
P(T_{ij}>t\mid \varepsilon_{ij}=l,X_{ij1},X_{ij2})=u_{ij}
\]
in $t$ to obtain
\[
1-\frac{F_l(t_{ij}\mid X_{ij1},X_{ij2})}
{F_l(\infty\mid X_{ij1},X_{ij2})}
=
u_{ij}.
\]
Equivalently,
\[
F_l(t_{ij}\mid X_{ij1},X_{ij2})
=
(1-u_{ij})F_l(\infty\mid X_{ij1},X_{ij2}).
\]
As for the proband, $t_{ij}$ was obtained by numerical inversion.

\item[{\it (iii)}] Compute
\[
Y_{ij}=\min(t_{ij},C_{ij}),
\qquad
\delta_{ij}=\mathbf{1}(t_{ij}<C_{ij})\times\varepsilon_{ij}.
\]
\end{enumerate}

\item[$\bullet$] For the proband, define similarly
\[
Y_{i1}=\min(t_{i0},a_i),
\qquad
\delta_{i0}=\mathbf{1}(t_{i0}<a_i)\times\varepsilon_{i0}.
\]
\end{enumerate}